\documentclass[12pt,preprint]{aastex}

\slugcomment{}

\shorttitle{Young Brown Dwarfs in the Core of the W3 Main}
\shortauthors{Ojha et al.}

\begin{document}

\title{Young Brown Dwarfs in the Core of the W3 Main Star-Forming Region}

\author{D. K. Ojha\altaffilmark{1}}
\affil{Tata Institute of Fundamental Research, Homi Bhabha Road, Colaba,
Mumbai 400 005, India}
\and
\author{M. Tamura, Y. Nakajima, H. Saito}
\affil{National Astronomical Observatory of Japan, Mitaka, Tokyo 181-8588, Japan}
\and
\author{A. K. Pandey}
\affil{Aryabhatta Research Institute of Observational Sciences, 
Manora Peak, Nainital 263 129, India}
\and
\author{S. K. Ghosh}
\affil{Tata Institute of Fundamental Research, Homi Bhabha Road, Colaba,
Mumbai 400 005, India}
\and
\author{K. Aoki}
\affil{Subaru Telescope, National Astronomical Observatory of Japan, 
650 North A'ohoku Place, Hilo, HI 96720}
\altaffiltext{1}{E-mail: ojha@tifr.res.in}

\begin{abstract}
We present the results of deep and high-resolution (FWHM $\sim$ 0\arcsec.35)
$JHK$ near-infrared observations with the Subaru telescope, to search 
for very low mass young stellar 
objects in the W3 Main star-forming region. The 
near-infrared survey covers an area of $\sim$ 2.6 arcmin$^2$ with 
10 $\sigma$ limiting magnitude exceeding 20 mag in the $JHK$ bands. The 
survey is sensitive enough to provide unprecedented details in W3 IRS 5 
and IRS 3a regions and reveals a census of the stellar population 
down to objects below the hydrogen-burning limit.
We construct $JHK$ color-color and $J-H/J$ and 
$H-K/K$ color-magnitude diagrams to identify very low luminosity young 
stellar objects and to estimate their masses. Based on these color-color and 
color-magnitude diagrams, we identified a rich population of embedded 
YSO candidates with infrared excesses (Class I and Class II), associated with 
the W3 Main region. A large number of red sources ($H-K >$ 2) have also 
been detected around W3 Main, which are arranged from the northwest 
toward the southeast regions. Most of these are concentrated around W3 IRS 5. We
argue that these red stars are most probably pre-main-sequence (PMS) stars with
intrinsic color excesses. 
We find that the slope of the $K$-band luminosity function of W3 Main
is lower than the typical values reported for young embedded clusters.
Based on the comparison between theoretical 
evolutionary models of very low-mass pre-main sequence objects with the 
observed color-magnitude diagram, we find there exists a substantial 
substellar population in the observed region. The mass function does not
show the presence of cutoff and sharp turnover 
around the substellar limit, at least at the
hydrogen-burning limit. 
Furthermore, the mass function
slope indicates that the number ratio of young brown dwarfs and 
hydrogen-burning 
stars in the W3 Main is probably higher than those in Trapezium and 
IC 348.
The presence of mass segregation, in the sense that relatively massive YSOs
lie near the cluster center, is seen. The estimated dynamical evolution
time indicates that the observed mass segregation in the W3 Main may be
the imprint of the star formation process.
\end{abstract}


\section{Introduction}

One of the most important problems in star formation is understanding 
the origin of the stellar initial mass function (IMF). The
pioneering work by Salpeter (1955) indicated that the IMF in the
mass range 0.4 - 10 M$_{\odot}$ can be presented with power law or 
lognormal functions. Recent studies have shown a break in the
function around 1 M$_{\odot}$ (cf. Kroupa 2001, 2002; 
Muench et al. 2002, 2003). It is believed that the majority of stars 
are formed in clusters which contain a large number of low mass stars. 
These low mass stars play an 
important role in the study of IMF. There are many embedded young 
clusters in molecular clouds (cf. Lada \& Lada 2003), which
are ideal targets for the study of IMF, because the young clusters are
known to keep their initial conditions of star formation. However, the
molecular clouds have long prevented optical observations of the embedded
young clusters, and we had to wait for the development of recent NIR arrays
for substantial observations (e.g., Lada \& Lada 2003). 

The functional form and the universality of the IMF at a very low mass 
regime including brown dwarfs (BDs) are still open questions. At the low end 
of the IMF are BDs which are star-like objects that span the mass range 
between giant planets like Jupiter and the least massive stars that are 
just below the hydrogen-burning limit but above the deuterium-burning 
limit. The expectation that BDs form by the same 
physical process as low-mass young stars suggests that one might find BDs in 
the very locations that one finds protostars: molecular clouds. However, it 
is not clear how abundantly BDs are formed in molecular clouds and whether 
they significantly contribute to the low-mass end of the luminosity 
function, and therefore to the IMF. By conducting surveys of the young 
stellar populations often associated with molecular cloud cores we will be 
able to characterize the IMF down into the BDs regime and determine: i) if 
there is a peak in the IMF of young, emerging star clusters near the 
hydrogen-burning limit (Zinnecker et al. 1993; Muench et al. 2002) 
and ii) what connection is there, if any, between 
such a peak and the physical conditions that characterize the molecular 
cloud. The main difficulty so far is the lack of high spatial resolution and 
sensitive data in the crowded fields, such as dense cluster cores, where 
there is a high probability of finding BDs.

So far the young brown dwarf (YBD) search 
has been limited to the nearby star forming regions ($<$ 500 pc; e.g. Orion, 
Taurus, Chamaeleon, Lupus, Ophiuchus: Kaifu et al. (2000), 
Lucas \& Roche (2000), Tamura et al. (1998), Oasa et al. (1999), 
Nakajima et al. (2000),
Comeron et al. (1993), Luhman \& Rieke (1999)).
In order to detect and characterize the YBDs in distant massive
star forming regions, {\it which are more typical in the galactic scale},
we need high sensitivity and high resolution.
Because the very low mass stars have infrared excesses when they are young, we 
can detect YSOs including YBDs only with the colors and magnitudes in the $JHK$ 
wavelengths (Muench et al. 2001; Liu et al. 2003; Luhman et al. 2007). 
Since stars are brighter when they are young especially at 
infrared wavelengths, it is easier to detect very low mass stars in young 
clusters. Hillenbrand (1997), Hillenbrand \& Carpenter (2000), and  
Luhman et al. (2000) observed a rich 
massive star forming region, the Orion star forming region. They obtained 
results that suggest the slope of IMF changes at 0.2 M$_{\odot}$ and its shape 
continuously changes towards BD regime.
However, the result is derived from only one star forming region. 
Therefore, it is important to find out if it is universal among other 
distant massive star forming regions.

W3 giant molecular cloud complex is a well-studied star-forming region 
located in the Perseus arm at a distance of 1.83 kpc that contains objects 
such as H II regions, embedded IR sources (including the extremely luminous 
cluster of sources W3 IRS 5), OH and water masers. In order to search for the 
embedded very low-mass young stellar populations, we have recently analysed a 
deep $JHK_{\rm s}$ (simultaneous) survey of W3 Main star-forming region 
(Ojha et al. 2004a). The observations were carried out with the SIRIUS camera 
on the UH 2.2 m telescope. The near-infrared (NIR) survey covers an area of 
$\sim$ 24 arcmin$^2$ with 10 $\sigma$ limiting magnitudes of $\sim$ 19.0, 
18.1, and 17.3 in $J$, $H$, and $K_{\rm s}$ bands, respectively. 
A significant number of very low-luminosity YSOs have been found 
in the core region of W3 Main based on their NIR colors (Ojha et al. 2004a), 
where the estimated 
stellar density is of $\sim$ 2000 pc$^{-3}$ (for $K$ $<$ 17.5 mag). We 
expect that some of them would be candidates for YBDs recently formed 
in the W3 molecular cloud. 

In this paper we present a new set of $J$-, $H$-, and $K$-band data 
for the W3 Main star-forming region with higher resolution and 
sensitivity ($K$ $\sim$ 20 mag)
for an area of $\sim$ 2.6 arcmin$^2$ centered on W3 IRS 5.
Our motivation
is to look for the YBDs associated with the W3 Main region and discuss
their nature and mass function (MF). 
In \S 2 and \S 3 we present the details of the observations and data
reduction procedures. Section 4 contains the results and discussion on
mostly point-like YSOs and describe the details of substellar population.
Section 5 compares the results from previous studies of W3 Main.
We then summarize our conclusions in \S 6.

\section{Observations}

Deep imaging observations of the W3 Main star-forming region at the NIR 
wavelengths $J$ ($\lambda$ = 1.25 $\mu$m), $H$ ($\lambda$ = 1.64 $\mu$m),
and $K$ ($\lambda$ = 2.21 $\mu$m) were obtained on 2004 August 27, using
the facility instrument Cooled Infrared Spectrograph and Camera for OHS 
(CISCO) mounted on the Cassegrain focus of the Subaru 8.2 m telescope. CISCO 
is equipped with a 1024 $\times$ 1024 Rockwell HgCdTe HAWAII array. A 
plate scale of 0\arcsec.105 pixel$^{-1}$ at the f/12 focus of the 
telescope 
provides a field-of-view of $\sim$ 1\arcmin.8 $\times$ 1\arcmin.8 
(Motohara et al. 2002). The off-target sky frames located at 
$\sim$ 30\arcmin~north of the target position were also taken just after the 
object frames. The sky frame was also used as a
reference field for W3 Main to assess the stellar
populations within the W3 Main star-forming region (see \S 4). 
The exposure times for individual frames were
40, 20, and 20 s in $J$, $H$, and $K$, respectively, yielding a total 
integration time of 12 minutes in all
three bands. Each position was observed in a 3 $\times$ 3 dithered pattern with
10\arcsec~offset, and three to six images were obtained at each dithered
position. All the observations were done under excellent photometric sky 
conditions. The average
seeing size was measured to be 0\arcsec.35 FWHM in all three filters.
The air mass varied only between 1.09 and 1.38. The United Kingdom Infrared
Telescope faint standard star, FS6 (Hawarden et al. 2001), was observed 
at air mass values close to the target observations, for the photometric 
calibration.     

\section{Data Reduction}

Data reduction was done using the Image Reduction and Analysis Facility 
(IRAF) software package\footnote{IRAF is distributed by the
National Optical Astronomy Observatory, which is operated by the
Association of Universities for Research in Astronomy, Inc., under contract
to the National Science Foundation.}. The sky flats were generated by
median-combining sky frames of the individual dithering sequences for each 
filter. 
The flat-fielding and sky subtraction with a median sky frame were applied.
Identification and photometry of point sources were performed by using the 
DAOFIND and DAOPHOT packages in IRAF, respectively. Because of source 
confusion and nebulosity within the region, photometry was obtained using 
the point spread function (PSF) algorithm ALLSTAR in the DAOPHOT package 
(Stetson 1987). We used an aperture radius of 4 pixels ($\sim$ 0\arcsec.4)
with appropriate aperture corrections per band for the final photometry.
We evaluated photometric accuracy by direct comparison of our results with
SIRIUS photometry (Ojha et al. 2004a). We selected point sources that are
measured in all three bands with errors $<$ 0.1 mag. Comparing between
our photometry and SIRIUS photometry, the sources with $J-H < 2$ 
have photometric dispersions of $\sim$ 0.10 mag in $JHK$ bands. The dispersion 
increases to $\sim$ 0.15 mag for the sources with redder colors
($J-H > 2$). Our higher spatial resolution permits better source
separation and sky determination. We have assumed that the resulting 
$J$, $H$, and $K$ magnitudes are the same between CIT system and the
SUBARU/CISCO system. Absolute
position calibration was achieved using the coordinates of a number of 
stars from the SIRIUS photometric data by Ojha et al. (2004a). The position
accuracy is better than $\pm$0\arcsec.04 rms in the W3 Main field. 
 
The completeness limits of the images were evaluated by adding artificial
stars of different magnitudes to the images and determining the fraction
of these stars recovered in each magnitude bin. The recovery rate was
greater than 90\% for magnitudes brighter than 20.7, 19.0, and 18.5 in the 
$J$, $H$, and $K$ bands, respectively. The observations are complete (100\%)
to the level of 18.5, 17.5, and 17.0 mag in the $J$, $H$, and $K$ bands, 
respectively. 
The 10 $\sigma$ limiting magnitudes for our observations are estimated
to be $\sim$ 22, 21, and 20 in the $J$-, $H$-, and $K$-bands, respectively.

We also assessed the photometric incompleteness of our sample around 
the central W3 IRS 5 cluster and diffuse region to the northeast 
(shown by a box drawn with the dashed lines in Figure 1) and compared it
with the incompleteness for the whole region. We do not find significant
difference in the completeness ratio as a function of magnitudes 
between the two regions. The completeness ratios 
for the cluster and northeast region are $\sim$ 10\% less as compared 
to the whole region at fainter magnitudes.
We therefore applied an averaged completeness factor across the 
field for further analysis, as it will not affect our qualitative
results. 

\section{Results and Discussion}

\subsection{Morphology}

The $J$-, $H$-, and $K$-band images of the W3 Main star-forming region are 
shown in Figure 1. A cluster of young stellar objects is seen around the
W3 IRS 5 source, which has been detected only in the $K$ band. 
The individual compact H II regions, ultracompact H II regions, and embedded
IR sources are marked in the $K$-band image. The density enhancement is
seen around IRS 5 molecular clump in the $K$-band.
A composite color image was contructed from the CISCO $J$-, $H$-, and 
$K$-band images ($J$ represented in blue, $H$ in green, and $K$ in red) and
is shown in Figure 2. This image is the highest resolution NIR image of the
region from the ground and probably the deepest to date. Our deep and 
high-resolution NIR images show 
distinctive reddish and bluish nebulosity features, dark filaments between
the diffuse nebulosity, and a significant population of faint stars in
W3 Main. A large number of red young stars that are presumably embedded 
in the molecular core are also seen around IRS 5 and IRS 3a.
The overall morphology and sizes of the bright nebulosities are 
quite similar to those in {\it Hubble Space Telescope (HST)} NIR images 
within $\sim$ 1\arcmin.0 $\times$ 1\arcmin.0 (Megeath et al. 2005). 
However, the comparison of the K-band images (see Appendix and 
Fig. 3 in Megeath et al. 2005) shows that the Subaru images looks a 
little deeper than the {\it HST} images, in spite of the their 
slightly better resolution.
Notable nebulosity features are summarized in Appendix.

\subsection{Photometric Analysis of Pointlike Sources}

We obtained photometric data of 295 sources within a 2.6 arcmin$^2$ 
area in $JHK$ bands with photometric 
errors in each band of less than 0.1 mag. 
Megeath et al. (2005) presented {\it HST} NICMOS imaging of W3 IRS 5, a binary
high-mass protostar. They showed three new 2.22 $\mu$m sources (NIR 3 - 5) 
with very red colors in addition to the two protostars (NIR 1 and NIR 2). 
NIR 1 and NIR 2 are saturated in the Subaru $K$-band image, whereas the
three new sources (NIR 3 - 5) are clearly resolved. 
In addition to this we find an additional
source about 0\arcsec.8 southwest of NIR 4. We designate this source 
as NIR 6. While comparing with the photometry, we find that NIR 6 is 
probably not an independent source but is most likely an unresolved gaseous
knot in the nebula (see Appendix and related Figure 14a for a more detailed 
discussion of the W3 IRS 5 system).   

We compared our derived photometry for the NIR 3 - 5 sources to those 
in Megeath et al. (2005). The agreement of magnitudes is 
encouraging except for IRS 3 which is a young OB star embedded in the
molecular core (Megeath et al. 2005). It is likely that NIR 3 has 
brightened by $\sim$ 0.6 mag within a period of above six years 
(1998 - 2004). 
In addition to NIR 1 - 6, we 
see several red sources and a detailed structure of the IR nebula just to 
the north of NIR 1 in our deep $K$-band image (see Appendix).

\subsubsection{Color-Color Diagram}

Figure 3 shows $JHK$ color-color (CC) diagrams of the W3 Main star-forming
region and the reference field, respectively, for the sources detected in 
the $JHK$ bands with photometric errors in each color of less than 0.1 mag. 
The reference field is also used for correction of field star 
contamination in the raw KLF of W3 Main (see \S 4.4). Due to
heavy extinction, only a limited number of sources have been detected in
$J$ band. The solid thin curve represents the locus of main-sequence (MS) and 
thick-dashed curve is for giant stars taken from Bessell \&  Brett (1988). 
The dotted line represents the classical T Tauri (CTT) locus as determined 
by  Meyer et al. (1997). 
Presuming universal nature of the extinction law in NIR
region (Mathis 1990), we plotted reddening vectors of the normal
reddening law as parallel dashed lines. We assumed that $A_J/A_V$ =
0.265, $A_H/A_V$ = 0.155, $A_K/A_V$  = 0.090 for CIT system (Cohen et
al. 1981). As can be seen in Figure 3, the stars in W3 Main are distributed
in a much wider range than those in the reference field, which indicates 
that a large fraction of the observed sources in W3 Main exhibit NIR 
excess emission characteristics of young stars with circumstellar material, 
as well as a wide range of reddening. 
We have classified the CC diagram into three regions (e.g., 
Sugitani et al. 2002; Ojha et al. 2004a, 2004b) to study the nature of 
sources. The ``F'' sources are located within the reddening band of the MS 
and giant stars. These stars are generally considered to be either field stars, 
Class III objects, or Class II objects with small NIR excess. ``T'' sources 
are located redward of region F, but blueward of the reddening line 
projected from the red end of the CTT locus of Meyer et al. (1997). These 
sources are considered to be mostly T Tauri stars (Class II objects) with 
large NIR excess. ``P'' sources are those located in the region redward of 
region T, and are most likely Class I objects (protostellar objects). There 
may be an overlap in the NIR colors of the upper end band of Herbig Ae/Be 
stars and in the lower end band of T Tauri stars in region T 
(Hillenbrand et al. 1992).
The majority of sources in our sample are distributed in the F region.
The total number of ``T'' (Class II) and ``P'' (Class I) sources are 89 and
78, respectively. However, this is the lower limit, as several cluster 
members detected in the $K$ band may not be detected in the other two 
shorter NIR wavelength bands. 

\subsubsection{Color-Magnitude Diagram}

Figure 4 ({\it left}) shows a $H-K$ versus $K$ 
color-magnitude (CM) diagram of 
the W3 Main star-forming region, where all the 
sources detected in the $JHK$ bands plus some 160 stars fainter than 
our limit at the $J$ band but still above the detection threshold in the
$H$ and $K$ bands are plotted.  
The vertical solid lines represent zero-age-main-sequence (ZAMS) 
curves (for a distance of 1.83 kpc) reddened by $A_V$ = 0, 15, 30, 45, 
and 60 mag, respectively. The parallel slanting lines represent the 
reddening vectors to the corresponding spectral type. An apparent MS track 
is noticeable at $H-K$ $\sim$ 0.4 in this diagram; 
however, a comparison of it with a similar diagram for the stars in the 
control field (Fig. 4; {\it right}) shows that 
it is a sequence caused by field stars in the foreground of W3 Main.
It also appears that the reference field has foreground extinction of the order 
of 3 - 5 visual magnitudes.
A comparison of the $J-H$ versus $H-K$ CC and $H-K$ versus $K$ CM diagrams 
of the reference field and W3 Main region indicates that
stars having $H-K < $ 1.0 are probable foreground sources. 
In Figure 4 ({\it left}), YSOs (Classes II and I) found from the CC diagram (Fig. 3) are
shown as stars and triangles, respectively. However, it is important to 
note that even those sources not shown with stars or triangles may also be
YSOs with an intrinsic color excess, since some of them are detected in
the $H$ and $K$ bands only and are not in the $J$ band because of their
very red colors. Figure 4 ({\it right}) shows that there is a lack 
of sources brighter than about $K$ = 14.5 in the reference field, which
are saturated in our observations. We have complemented the control
field observations with 2MASS sources which are shown with cross symbols.
The number of such bright stars in our 
field-of-view is small and does not contribute the object field as extincted 
background stars that mimic YSOs.

\subsection{Spatial Distribution of Young Stellar Sources}

In our deep NIR observations, 130 very red sources are detected in
the $H$ and $K$ bands, out of which only 10 sources have $J$
counterpart. The sources having colors redder than $H-K > 2$ in
Figure 4 ({\it left}) are also considered as YSO candidates. 
In Figure 5 the spatial
distribution of the YSO candidates identified in Figures 3 and 4
is shown. Stars represent sources of T Tauri type (Class II),
filled triangles indicate Class I sources, and filled circles
denote the very red sources ($H-K > 2$). The white contours in
Figure 5 represent $^{12}$CO(1--0) emission obtained with Nobeyama
Radio Observatory 45 m telescope (T. Sakai, private communication).

In general, Class I and Class II candidates are distributed all
over the field; however, an apparent concentration of
Class I sources (filled triangles) around W3A, W3B, and
$\sim$ 30\arcsec~southeast of W3A H II regions can be noticed in Figure 5. 
It appears that most of these YSOs (Class I sources) are associated with
the diffuse ionized gas at the edge of these H II regions. Most of
the sources with large color indices ($H-K > 2$) also seem to
be distributed all over the field, barring the W3A H II region. However,
an apparent concentration of these sources can be seen around W3F, IRS 7
and IRS 5 regions (see Figs. 1 \& 5). The lack of these red sources around
W3A H II region indicates that this region is relatively evolved as 
compared to other regions. 

If we assume that the large $H-K$ ($>$ 2) color results merely from
the interstellar reddening affecting normal stars, then the
extinction value might even exceed 60 mag in the molecular cloud
in which some of the red stars are found (see Fig. 4). However,
with such a large amount of absorption, diffuse emissions are
unlikely to be detected in the NIR. Since most of the red sources
are associated with faint diffuse emission across the field (Fig. 5),
this provides evidence that these sources are YSOs with intrinsic
NIR excesses and possibly local extinction as well. We also find 
that the mean $A_V$ value of the Class II sources around 
W3 IRS 5 dense cluster region is about 21 mag. 
A similar value of $A_V$ for the red sources indicates that large
$H-K$ color of these sources should be due to NIR excess. 
This further provides evidence
that the red sources should be PMS objects still embedded in clouds.

\subsection{The $K$-band Luminosity Function}

The $K$-band luminosity function (KLF) of an embedded cluster is 
useful in constraining the age of the cluster (Megeath et al 1996; 
Ojha et al 2004a). Our motivation is also to compare the KLF slope
obtained from deeper Subaru $K$-band data with the results from 
previous studies of W3 Main (Ojha et al. 2004a; Megeath et al. 1996). 
In order to derive the observed KLF, one needs to apply corrections for 
(1) the incompleteness of the star counts as a function of the $K$ 
magnitude, and (2) the field star contribution in the line of sight of the 
cluster.

We have determined the completeness of our $K$-band data through artificial
star experiments using ADDSTAR in IRAF (see \S 3). This was performed 
by adding fake 
stars in random positions into the images at 0.5 magnitude intervals and 
then by checking how many of the added stars could be
recovered at various magnitude intervals. We repeated this procedure
at least five times. We thus obtained the detection rate as a function of
magnitude, which is defined as the ratio of the number of recovered
artificial stars over the number of added stars.

One method to correct for the foreground and background contamination is
to use a reference field, located reasonably away from the cluster, and
yet close enough to have a similar Galactic field star distribution, e.g.,
with the same Galactic latitude. The observed reference field is then
assumed to represent the contaminating field population in the W3 Main
region. However, in the case of a young and partly embedded region like 
W3 Main, the colors and magnitudes of the background contaminating 
populations are significantly affected by the parent molecular cloud.
To correct for the foreground and background contamination, we 
used of both the Galactic model by 
Robin et al. (2003) and the reference field star counts. This Galactic model 
of population synthesis reproduces the stellar contents of the Galaxy.
The star counts were predicted using the Besan\c con model of 
stellar population synthesis (Robin et al. 2003) in the direction of the 
reference field close to the W3 Main (see \S 2), which is also corrected 
for photometric completeness. 
To ensure that the model does a reasonably good job in predicting the field
star population, we compared the model star counts with those of our observed
reference field. Figure 6 shows the comparison of the completeness corrected
KLF of the reference field with the model star counts where all the stars
are made fainter by an interstellar extinction of $A_V$ = 5 mag.
It can be seen that the model star counts match reasonably well
with the observed reference field. Hence, we used this model to predict
the contamination to the W3 Main field, for which all the field stars
are made fainter by an interstellar extinction of $A_V$ = 5 mag, 
whereas the background stars are made fainter by an additional cloud
extinction.

The advantage in using this model is that the background stars 
($d >$ 1.83 kpc) can be separated from the foreground stars ($d <$ 1.83 kpc).
While all the stars in the field suffer a general interstellar extinction, 
only the background stars suffer an additional extinction due to the 
molecular cloud. From Figure 4 we conclude that
the average extinction to the embedded cluster is $A_V$ $\sim$ 15 mag
($H-K$ $\sim$ 1.0). Assuming spherical geometry, background stars are seen 
through $A_V$ = 30 mag (2$ \times$ 15 mag). Therefore, we applied an 
extinction value of $A_V$ = 30 mag (or $A_K$ = 2.7 mag) in simulating the 
background stars. We combined the foreground and background stars to make a 
whole set of the contamination field and obtained the fraction of the 
contaminating stars over the total model counts. Then we scaled the model 
prediction to the star counts in the reference field, and subtracted the 
combined foreground ($d <$ 1.83 kpc) and background ($d >$ 1.83 kpc with 
$A_K$ = 2.7 mag) data from the KLF of the W3 Main region.

After correcting for the foreground and background star contamination and
photometric completeness, the resulting KLF is presented in Figure 7 for 
the W3 Main region, which follows a power-law in shape. In Figure 7, a 
power-law with a slope $\alpha$ [$d N(m_K)/dm_K \propto 10^{\alpha m_K}$, 
where $N(m_K)$ is the number of stars brighter than $m_K$] has been fitted 
to the KLF using a linear least-squares fitting routine. The KLF of the 
W3 Main region shows a power-law slope of $\alpha$ = 0.14 $\pm$ 0.03
over the magnitude range 14 - 18 in the $K$-band.
The derived KLF slope is lower than those 
generally reported for the young embedded
clusters ($\alpha \sim 0.4$, e.g., Lada et al. 1991, 1993; Lada \& Lada 2003).
Thus, this low value of the slope is indeed an intrinsic property of the
stellar population in the W3 IRS 5 region.
 
\subsection{Age and Stellar Mass Estimates}

The fraction of NIR excess stars in a cluster is an age indicator because 
the disks/envelopes become optically thin with the increasing age. For young 
embedded clusters having age $\sim 1 \times 10^6$ yr, the fraction of NIR 
excess stars (based on $JHK$) is found to be  $\sim$ 50\% (Lada et al. 2000; 
Haisch et al. 2000) to $\sim$ 65\% (Muench et al. 2001).
The fraction, in the case of Taurus dark clouds (age $\sim 1-2 \times 10^6$ yr)
is estimated as 40\% (Kenyon \& Hartmann 1995), which decreases to 
$\sim$ 20\% for 
the clusters with age $\sim 2-3 \times 10^6$ yr (Haisch et al. 2001; 
Teixeira et al. 2004; Oliveira et al. 2005).

To estimate the fraction of the NIR excess stars in our sample,
we have to estimate the contamination in W3 Main region due to field stars.
In order to estimate the foreground and background contaminations, we 
made use of both the observed star counts and the Galactic model 
(see more details in \S 4.4). Using the model predictions and the 
reference star counts, we obtained a fraction of about 17\% foreground
and background contaminating stars in our sample.  
After correcting for the photometric completeness, foreground and
background star contamination, the fraction of the NIR excess stars can
be estimated to be 65\%, 
which suggests an upper
age limit of $\sim 1 \times 10^6$ yr for the W3 Main region. However, to
further improve the statistics, observations in $L$ and $M$-bands are 
needed.

In their NIR survey of the central region 
($\sim$ 1\arcmin.5 $\times$ 1\arcmin.5) around W3 IRS 5, Megeath et al. (1996) 
found that about 30\% of the NIR sources were Class I objects and
concluded that the age of the low mass population is $> 0.5$ Myr. 
Present data also indicate that about 34\% YSOs are the Class I sources. 
On the basis of $HST$ NICMOS imaging of W3 IRS 5 region, 
Megeath et al. (2005) proposed a nascent Trapezium system in the center of 
the W3 IRS 5 cluster. On the basis of above discussions we assume an age 
of 1 Myr for the W3 IRS 5 cluster region.

We estimated the extragalactic contamination in the NIR excess sources
of our sample (see also Oasa et al. 2006) by assessing the number of
extincted galaxies (or faint background galaxies) using an average
number of galaxy counts (Gardner et al. 1993; McLeod et al. 1995;
Saracco et al. 1999; Maihara et al. 2001). We assume an average extinction
of $A_V$ = 30 mag (or $A_K$ = 2.7 mag; see \S 4.4) due to the W3 Main 
region. The total
number of galaxies in our observed region, i.e. $\sim$ 2.6 arcmin$^2$,
is predicted to be $\sim$ 3 down to our 10 $\sigma$ limiting extincted magnitude
of 20 in the $K$-band. We therefore conclude that the NIR excess
sources in our sample are most likely YSOs associated with the molecular cloud.

The mass of the probable YSO candidates can be estimated by comparing 
their locations on the CM diagram with evolutionary models of PMS stars. 
To minimize the effect of the ``excess'' emission in the NIR, we used 
($J-H$) versus $J$ CM diagram instead of the ($H-K$) versus $K$ 
CM diagram to derive the ages and masses of the YSOs. Figure 8 represents 
($J-H$) versus $J$ CM diagram for probable YSOs (Classes II and I). The 
symbols are same as in Figure 4. 
The solid curve in Figure 8 denotes the loci of 1 Myr old PMS stars 
from Palla \& Stahler (1999) over the mass range 1.4 - 4.0 $M_{\rm \odot}$, at 
an assumed distance of 1.83 kpc. The dotted curve represents a PMS isochrone 
of 1 Myr from Baraffe et al. (1998; 2003) for low-mass YSOs (0.005 - 1.4 
$M_{\rm \odot}$). The solid and dotted oblique reddening lines denote 
the positions of PMS stars of 2.0, 0.1, 0.02, and 0.01 $M_{\rm \odot}$ for
1 Myr. Most of the objects well above the PMS tracks are luminous
and massive ZAMS stars (see Table 1 in Ojha et al. 2004a). 
As can be seen in Figure 8, the majority of YSOs have masses in the 
range of 0.01 - 2.0 $M_{\rm \odot}$, indicating a significant population 
of very low mass stars below the hydrogen-burning limit.

\subsection {Mass Function}

The interpretation of the KLF as IMF is complicated because of the
NIR excess arising from the circumstellar disks. We therefore prefer 
$J$-band luminosity function (JLF) to determine the MF of the YSOs 
(Class II \& Class I), identified using their NIR excess properties,
which is least affected by the circumstellar matter compared to what is expected
solely from the stellar photosphere. We assume that all the YSOs
are members of the W3 Main field and the field contamination is negligible 
in the line of sight of the cluster (see Fig. 4 \& \S 4.2.2). 
We constructed reddening-corrected JLF for
the Class II sources (Fig. 9; {\it left}). Their luminosities are corrected for 
individual extinction determined by dereddening to the reddening-free 
locus of classical T Tauri stars or its extension in ($H-K$) versus
($J-H$) CC diagram (Fig. 3) as determined by Meyer et al. (1997). 
The JLF was also corrected for completeness which was evaluated by
using artificial star experiment (see \S 3). 
We also constructed JLF for both the Class II and Class I sources, although 
Class I objects are much younger and therefore more luminous than 
Class II objects. We assumed an average extinction (derived from the 
individual extinction estimates for the identified Class II objects)  
for all the Class I objects and dereddened these objects 
with this average extinction value. Figure 9 ({\it right}) presents the 
JLF for Class II and Class I sources. 
The completeness limits used in the JLFs (Fig. 9) are
determined from the dereddening of the $J$-band completeness limits 
(see \S 3) by an average extinction.  
By looking at the star - brown dwarf (hereafter star-BD) boundary
at 1 Myr in Figure 9, there 
appears to be a substantial substellar population in W3 Main and YBDs 
can thus be distinguished from low mass ($\ge$ 0.1 $M_{\odot}$) YSOs 
in this JLF.    

We derived a relation between $J$-band luminosity and mass for evolutionary
models of very low mass objects from Baraffe et al. (1998), 
Baraffe et al. (2003), and Palla \& Stahler (1999). The Baraffe et al.'s
models are designed for very low mass objects, including giant planets, while
the Palla \& Stahler (1999) model covers a relatively intermediate mass
range. Figure 10 shows the derived mass - $J$-band luminosity
relationship for the 1 Myr age based on the theoretical models of
Baraffe et al. (1998; 2003), and Palla \& Stahler (1999).
We find that the derived isochrones are in reasonable 
agreement for $\geq$ 0.01 $M{_\odot}$. 
Using this mass-luminosity relation, the star-BD boundary 
(0.075 $M_{\odot}$) at 1 Myr corresponds to $J_0$ $\sim$ 17.75 mag.  

The JLF was converted to MF using the mass - $J$-band luminosity relation.
The MF of Class II objects was obtained by counting the number of stars 
in various mass bins and is shown in Figure 11 ({\it left}). We have 
constructed MF in wide bins in order to minimize uncertainties in the 
mass estimate. 
Figure 11 ({\it right}) also presents the composite MF for Class II and
Class I sources. Overall, the shape of the composite MF looks similar
to the Class II sources (Fig. 11), but with a different slope.

The MF of W3 Main (Fig. 11) appears to rise monotonically
across the hydrogen-burning limit up to $\sim$ 0.04 $M{_\odot}$ and
seems to have resemblance with that of the S106 (Oasa et al. 2006), whereas 
the MF for the center core region 
in the ONC with the detection limit $\sim$ 0.02 $M{_\odot}$, is characterized
by a local peak near 0.2 - 0.3 $M{_\odot}$ and appears to fall into the BD
regime (Lucas \& Roche 2000; Hillenbrand \& Carpenter 2000). A similar
trend is seen by Muench et al. (2002) in their study of MF in
the Trapezium cluster.
It is to be noted that at $\sim$ 0.1 $M{_\odot}$, the completeness
factor of our $J$-band data is about 84\% and it falls to 50\% around 
0.05 $M{_\odot}$.
The MFs have slopes ($dLog (N)/dLog(M)$) of -0.23 $\pm$ 0.02 and
-0.66 $\pm$ 0.04, respectively, for the Class II and composite samples, 
over the mass range 0.04 $<$ $M/M_{\rm \odot}$ $<$ 1.6. 
Here, it is worthwhile to mention that Class I sources have higher NIR 
excess, therefore the mass estimates of these objects on the basis of 
JLF may be biased towards higher values. If it is true,
the slope of the MF for Class I objects will be steeper. 
We assess the issue of the increase of dispersion at fainter 
$J$ magnitudes which affects the de-reddening and hence MFs. We find 
that the dispersion is small enough to consider it unimportant as it 
does not change the overall shape of the MFs and therefore
the qualitative conclusions as drawn above are not affected.    

We also estimated the MFs derived using an average 
extinction for both the Class II and composite samples. 
The completeness-corrected MFs are shown as continuous thin lines in
Figure 11. The overall shapes of the MFs look similar and still indicate 
a monotonic rise across the hydrogen-burning limit up to 
$\sim$ 0.04 $M{_\odot}$. The qualitative results do not affect the 
conclusions as drawn above. However, the slopes of MFs have changed to 
-0.74$\pm$0.11 and -1.00$\pm$0.08, respectively, for the Class II and 
composite samples, over the mass range 
0.04 $<$ $M/M_{\rm \odot}$ $<$ 1.6. 

We also checked the distribution of MF by excluding the sources
in the ``P'' region that fall below $J-H$ = 1.0. It may be possible
that sources fall in this region of the diagram because of photometric 
defects. Figure 12 ({\it left}) presents the MF for the Class I and 
Class II sources by removing those sources that fall below $J-H$ = 1.0 
in the ``P'' region. Since there are very few such sources (12 sources) 
in our sample, the overall statistics do not change. The overall shape 
of the MF does not change and hence the conclusions drawn above are 
not affected.  
We further checked the effect of $A_V$ on the MF with the help of 
extinction distribution in the W3 Main region. This technique has been 
used in young galactic clusters which have variable amount of reddening 
(e.g. Pandey et al. 2008). To map the extinction in the W3 Main region 
we used sources lying in the ``F'' region, which show a range in the 
$A_V$ from 0 to 35 mag (see Fig. 3). To estimate the $A_V$, we traced 
the stars in ``F'' region in the reddening band to the solid line shown 
in the Figure 3, near the MS locus (M0 - M6). The whole region was 
divided into small cells of 200 $\times$ 200 pixels 
($\sim$ 0\arcmin.35 $\times$ 0\arcmin.35) so that each cell has 
significant number of stars. The mean $A_V$ value of each cell was 
used to deredden the Class II (``T'' region) and Class I (``P'' region)
sources lying in that cell. Figure 12 ({\it right}) shows the composite MF
for all the Class II and Class I sources lying in ``T'' and ``P''
regions. The completeness-corrected MF as 
shown by a dashed-dotted line in Figure 12 ({\it right}) still indicates a monotonic 
rise across the hydrogen-burning limit and the qualitative results 
do not affect the conclusions drawn in this section.

Comparison of MF of young star forming regions provides an important 
diagnostic to study the evolution of star formation process in these regions. 
Comparison can reveal similarities or differences in MFs that would test 
the universality of the MF. Comparison of the MF of the W3 Main and that of 
S106 by Oasa et al. (2006) reveals resemblance, in the sense that there is no
break in MF slope in BD regime. Whereas the MF for Trapezium and IC 348 
clusters declines below $0.12 M_\odot$ and  $0.08 M_\odot$, respectively 
(Muench et al. 2002; 2003).
The derived MF slopes indicate that the number ratio of YBDs and
hydrogen-burning stars in the W3 Main is probably higher than those in the 
ONC and its Trapezium cluster.

\subsection{Mass Segregation and the Formation Scenario of YBDs in W3 Main}

The distribution of YSOs can be used to probe the star formation history 
of the region. Figure 13 shows the spatial distribution of Class II objects, 
on the $K$-band image. In this Figure the blue circles show the sources 
with $J_{\rm 0}$ $<$ 17.25 mag ($M$ $>$ 0.1 $M_{\rm \odot}$) and the red 
ones are those with $J_{\rm 0}$ $>$ 17.25 mag ($M$ $<$ 0.1 $M_{\rm \odot}$). 
The separation at $J_{\rm 0}$ =  17.25 mag ($M$ $\sim$ 0.1 $M_{\rm \odot}$) 
is made as it allows us to distinguish relatively massive YSOs 
($J_{\rm 0}$ $<$ 17.25 mag) from possible YBDs ($J_{\rm 0}$ $>$ 17.25 mag)
in our sample.
It is noteworthy to see that most of the brighter ones ($J_{\rm 0}$ $<$ 17.25 mag; 
relatively massive YSOs) are concentrated at the center of the image 
around W3 IRS 5 and W3 IRS 3a regions, whereas the fainter ones 
($J_{\rm 0}$ $>$ 17.25 mag; possible YBD candidates) are away from the dense molecular 
regions. The distribution of other point sources appears to be quite 
uniform for both the $J_{\rm 0}$ mag ranges.

This result can be interpreted as follows: 1) This is a kind of
mass segregation where massive stars tend to stay in the cluster center
while low-mass stars are further away.
To characterize further the degree of mass segregation in W3 Main we show
the distribution of Class II sources, combined distribution of 
Class II + Class I 
sources, and all the sources lying in ``F'' region of the $(J-H)$ versus $(H-K)$
CM diagram (Fig. 3), as a function of radius in two different magnitude groups 
in Figure 14. The radial distance is calculated 
with respect to the coordinates of W3 IRS 5. The figure clearly indicates 
an effect of mass segregation in the radial distribution of Class II sources 
in the sense that the relatively massive YSOs 
(J$_{\rm 0}$ $<$ 17.25; $M$ $>$ 0.1 $M_{\rm \odot}$) tend to lie toward 
the cluster center, whereas possible YBDs 
(J$_{\rm 0}$ $>$ 17.25; $M$ $<$ 0.1 $M_{\rm \odot}$)
are found away from the center. The segregation is less prominent in the case
of combined distribution of Class I and Class II objects. Since we have applied 
a mean value of $A_V$ (see \S 4.6) to calculate the $J_{\rm 0}$ for Class I 
sources, this could be a probable cause for the dilution of segregation 
effect. The ``F'' region sources do not show any segregation;
although some patchy extinction just around W3 IRS 5 provides
an impression of a paucity of the ``F'' region sources near IRS 5,
the radial averaged $A_V$ is quite smooth out up to 0.6 pc.
Similar mass segregation has been reported in several 
star-forming regions, e.g., ONC, IC 348, S106 (cf. Oasa et al. 2006) as well 
as in young star clusters (e.g., Pandey et al. 2005; 
Sharma et al. 2007; Jose et al. 2008).
 
Dynamical effects could be the probable cause of the mass segregation.
Because of the dynamical relaxation, low mass stars in a cluster may acquire 
larger random velocities, consequently occupy a larger volume than high mass 
stars (cf. Mathieu 1985; Mathieu \& Latham 1986; McNamara \& Sekiguchi 1986). 
We estimated the relaxation time to decide whether the mass segregation 
discussed above is primordial or due to dynamical relaxation. To estimate the 
dynamical relaxation time $T_E$, we used the relation

\begin{displaymath}
T_{E} =  \frac{8.9\times10^5 N^{1/2} R_h^{3/2}}{\bar{m}^{1/2} log (0.4N)}
\end{displaymath}

where N is the number of cluster stars, $ R_{h}$ is the radius containing half of
the cluster mass and $\bar{m}$ is the average mass of cluster stars (Spitzer \& Hart
1971).
The lower limit of the stellar mass associated with the W3 IRS 5 system was 
estimated by Megeath et al. (2005) as about 54 $M_{\rm \odot}$. The total number of 
probable members of the cluster are estimated to be 170 
(3 main-sequence stars, 89 Class II objects, and 78 Class I objects).
Fifty percent of the probable members are found to be located within 0.6 arcmin 
($\sim$ 0.32 pc) of IRS 5. We assumed this value as the half-mass radius of the 
cluster.
Using these numbers, the estimated relaxation time $T_{E}$ comes out to be 
$\sim 2$ Myr, which, within uncertainity, is comparable to the age of the 
PMS stars of W3 Main. Here it is interesting to mention that Feigelson \& Townsley 
(2008) have found that some of the OB stars in the core of W3 Main are formed after 
the bulk of the more widely distributed cluster PMS stars. This form of age spread 
has long been noted in star clusters (e.g. Herbst \& Miller 1982;
DeGioia-Eastwood et al. 2001; Pandey et al. 2005, 2008; Sharma et al. 2007; 
Jose et al., 2008). If star formation is still going on in the region, 
the W3 Main OB stars confirms the extreme youth of the cluster, 
which consequently indicates that the observed mass segregation in the 
region may be of primordial nature.

2) Similar to 1) but alternatively, it might reflect the formation
scenario for brown dwarfs. If the ejection scenario for brown dwarf
formation is correct, we can also expect the same effect (i.e., BDs tend
to go away from the cluster center after their formation) and finally
3) it could also be due to some observation bias. The sensitivity of 
lowest mass objects could be slightly lower due to the higher stellar 
density toward the center or due to their associated nebulosity. 
This could be a problem in S106 where the nebula is brightest near
the center of the system. But at 
least the nebulosity around W3 IRS 5 is not a problem because we have 
detected YBDs even in the bright nebulosity around W3 A (Fig. 13). 

\section{Comparison With Previous Studies}

W3 Main region has been studied in detail at radio, millimeter, and infrared 
wavelengths, however, little is known about its low-mass stellar population.
Two previous studies (Megeath et al. 1996; Ojha et al. 2004a) report 
sources with $K$-band excesses indicative of Class I/II PMS stars. It is 
important to see how the new observations compare and extend the results
from previous studies. We find that the derived KLF power-law slope is in 
remarkable agreement with that of Megeath et al. (1996) and  
Ojha et al. (2004a) for the W3 IRS 5 cluster region, inspite of a very
different sensitivity and resolution. The observed cluster density
around W3 IRS 5 (within an apparent radius of 0.27 pc) is about 
$\sim$ 3000 stars pc$^{-3}$ for $K$ $<$ 20.5 as compared to 
$\sim$ 2000 stars pc$^{-3}$ for $K$ $<$ 17.5 (Ojha et al. 2004a).
This high density of stars in W3 Main may be a result of the youth 
of the cluster and the enormous mass of molecular gas available
in the W3 Main molecular core (Megeath et al. 1996).
Megeath et al. (1996) found that no more than 30\% of the NIR sources 
were Class I in the W3 Main region. The present deeper NIR survey also
indicates that about 34\% NIR excess sources are Class I. This implies 
that most of the PMS stars in W3 Main are Class III, as in most other 
young stellar clusters observed with {\it Chandra}, and the age of the 
low-mass population is $>$ 0.5 Myr (Feigelson \& Townsley, 2008). Ojha
et al. (2004a) estimated that the lowest mass limit of Class II and 
Class I candidates in their observations was 0.1 $M_{\rm \odot}$, whereas 
the high resolution deep imaging presented in this paper is essential for 
the census of brown dwarf IMF in star forming regions.

\section{Conclusions}

We have successfully conducted the deepest NIR imaging survey of 
the W3 Main star-forming region including IRS 5 with the highest 
spatial resolution in the star-forming regions thus far. Our main
results are summarized as follows:

1. Based on the $J-H$ versus $H-K$ diagram, approximately 167 YSO
candidates (Classes I and II) have been identified in a region of 
$\sim$ 1\arcmin.6 $\times$ 1\arcmin.6, with 10 $\sigma$ limiting 
magnitude more than 20 in the JHK bands.

2. The KLF of the W3 Main region around IRS 5 shows a power-law slope of
\mbox{$\alpha$ = 0.14 $\pm$ 0.03}, which is in good agreement with
the results of Megeath et al. (1996) and Ojha et al. (2004a).

3. We use the presently available theoretical mass-luminosity relation 
to convert the $J$-band luminosity function into mass function. There 
appears to be a substantial substellar population of YSOs 
based on the mass estimates on the basis of this mass -- dereddened
$J$-band luminosity relationship.

4. The distribution of bright and faint young stellar candidates
{\it versus} radial distance is suggestive of mass segregation effects. 
Extreme youth and estimated 
dynamical relaxation time indicate that the observed mass segregation 
in the W3 Main may be the imprint of star formation process.

5. According to our results, it is unlikely that the mass function shows  
the presence of cutoff and a sharp turnover around the substellar limit, at 
least at the hydrogen-burning limit.

6. If the center of the W3 IRS 5 cluster is a nascent Trapezium system 
(Megeath et al. 2005) then the number ratio of BDs and hydrogen-burning 
stars in the W3 Main is probably higher than that in the Trapezium 
cluster. 

7. If confirmed, these BDs will be the lowest mass members of the 
W3 IRS5 cluster and therefore provide the key clues to a census of very 
low mass stars and the IMF down down to about 30 Jupiter-mass for the first
time in distant massive star forming region.

\acknowledgments
We thank the anonymous referee for a critical reading of
the paper and several useful comments and suggestions, which 
greatly improved the scientific content of the paper.
This research made use of data collected at Subaru Telescope, which is 
operated by the National Astronomical Observatory of Japan. We are 
grateful to the Subaru Telescope staff for their support. 
We thank T. Sakai for providing us with the FITS image of his CO 
molecular line observation of W3 Main. We thank Annie Robin for 
letting us use her model of stellar population synthesis.
D. K. O. was
supported by the Japan Society for the Promotion of Science through a 
fellowship, during which most of this work was done.

\appendix

\section{Selected Notable Features}

In Figure 15 we present selected areas of the W3 Main star-forming
region in the deepest NIR images so far that are of noteworthy interest.

1. Figure 15a shows the resolved high-mass binary protostellar system
W3 IRS 5. A compact nebula is seen between the NIR 1 and NIR 2. The nebula 
may result from the collision of outflows from NIR 1 and/or NIR 2 
(Megeath et al. 2005). We see a dense cluster of embedded stars around IRS 5, 
a number of these red stars are presumably embedded in the molecular core 
(Fig. 5).  

The previous mid-infrared (MIR) images in fact show the presence of 4 sources 
of which MIR 1, 2, and 3 are described in van der Tak, Dianchi, \& Tuthill
(2005). If we
compare their 4.7 $\mu$m image with our $K$-band Image (Fig. 15a), it 
appears that MIR 2 is in fact a double source and corresponds well with 
NIR 1 and NIR 2, respectively. But if we compare with longer wavelength 
($>$ 9.9 $\mu$m) images, the separation between MIR 2 and MIR 1 gets 
slightly (0\arcsec.1 - 0\arcsec.2) shorter (in fact their 9.9 $\mu$m 
MIR 2 peak is extended), and MIR 2 does not exactly coincide with NIR 2 but 
within the ``bridge'' of nebulosity.

The radio source C or K5 (Claussen et al. 1994; van der Tak, Dianchi, 
\& Tuthilli, 2005)
appears to coincide with the nebula associated with NIR 1 to the west,
whereas NIR 5 is marginally associated with an unnamed radio source of 
Claussen et al. (1994). There is no NIR ounterpart to radio source a
(Claussen et al. 1994).

There is a funny shaped red nebula to the northwest of NIR 1 (Figs. 2 and
15a). This is apparently connected with NIR 1. Interestingly, the nebulosity
around this nebula has not been mentioned earlier. Its nature is 
unknown and no radio source is associated with this nebula 
(Claussen et al. 1994). However, the direction of the extension
of this red nebula is consistent with the large-scale
outflow direction (Fig. A1 in Gibb et al. 2007). Thus, this nebula could 
be a reflection nebula associated with the outflow. The red-shifted 
component (to southeast) might be invisible even at $K$-band
due to much heavier extinction.

2. Figure 15b shows the nebulosity associated with the compact H II 
region W3 B excited by a zero-age main-sequence O-type star IRS 3a. 
Equally noticeable in Figure 15b are the dark lanes which
bisect the nebula. Prominent ones are on the northeast side of the 
ionized gas and thus are regions of increased extinction comprising
ambient molecular gas that has not yet been ionized by the UV radiation
field. The appearance of this nebula (Fig. 15b) is very similar to the
famous Trifid Nebula (M 20, NGC 6514) in Sagittarius excited by an 
O star.
 
3. Dark filamentary lanes are seen (Fig. 15c), with irregular shapes extending 
throughout the whole W3 Main region. They are associated with the dense
molecular gas (Fig. 5). An infrared source (detected only in the $H$ and $K$-
bands) marked by an arrow with large color excess ($H-K$ = 4.48) is located 
inside the dark lanes. This is most
probably a young star in its earliest evolutionary phase. 

4. We detect a faint nebulosity around the ultracompact H II region W3 E
(Fig. 15d). Dark edge adjacent to the bright nebulosity is particularly
striking.

\clearpage

\begin{table}
\caption{NIR sources around W3 IRS 5}
\begin{tabular}{ccccc}
\\
\tableline
\tableline
Source  & R.A.      & Decl.     & $HST$ $F222M$\tablenotemark{\dagger} & Subaru $K$-band \\
        & (J2000.0) & (J2000.0) &    (mag)      &    (mag) \\
\tableline
NIR 3 & 2 25 40.73 & +62 05 49.9 & 17.94 $\pm$ 0.10 & 17.38 $\pm$ 0.04\\ 
NIR 4 & 2 25 40.90 & +62 05 51.7 & 16.20 $\pm$ 0.07 & 16.09 $\pm$ 0.07\\
NIR 5 & 2 25 41.04 & +62 05 52.1 & 17.94 $\pm$ 0.10 & 17.90 $\pm$ 0.07\\
\tableline
\end{tabular}

\tablenotetext{\dagger}{Megeath et al. (2005)}
\end{table}

\clearpage

\begin{figure}
\epsscale{0.8}
\plotone{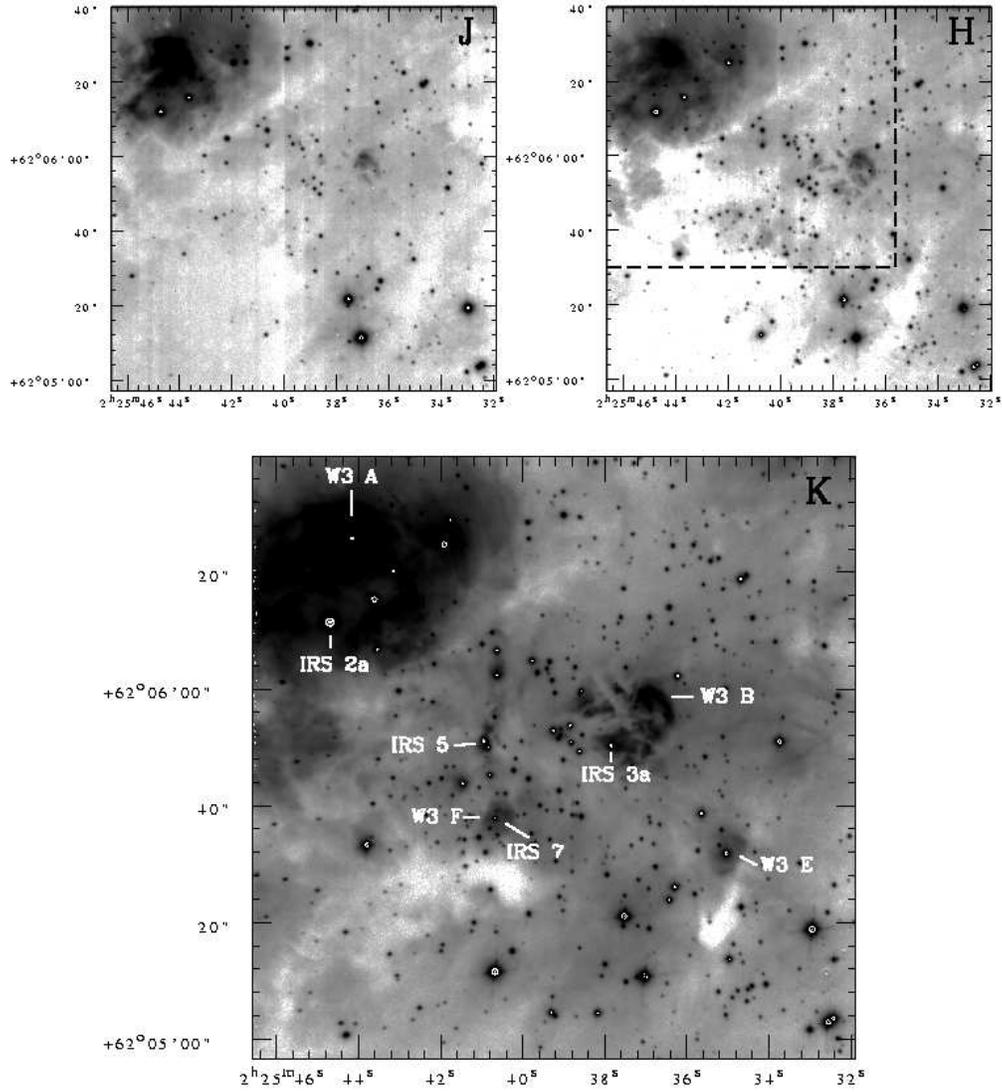}
\caption{$J$-, $H$-, and $K$-band images of the W3 Main star-forming region  
displayed in a logarithmic intensity scale. The box marked with the 
dashed lines in the $H$-band image shows the cluster region around 
W3 IRS 5 and diffuse region to the northeast.
The locations of the individual H II and ultracompact H II regions and 
the embedded IR sources are marked in the $K$-band image. 
North is up and east is to the left. The abscissa and the ordinate are in 
J2000.0 epoch.
\label{fig1}}
\end{figure}

\clearpage

\begin{figure}
\epsscale{1}
\plotone{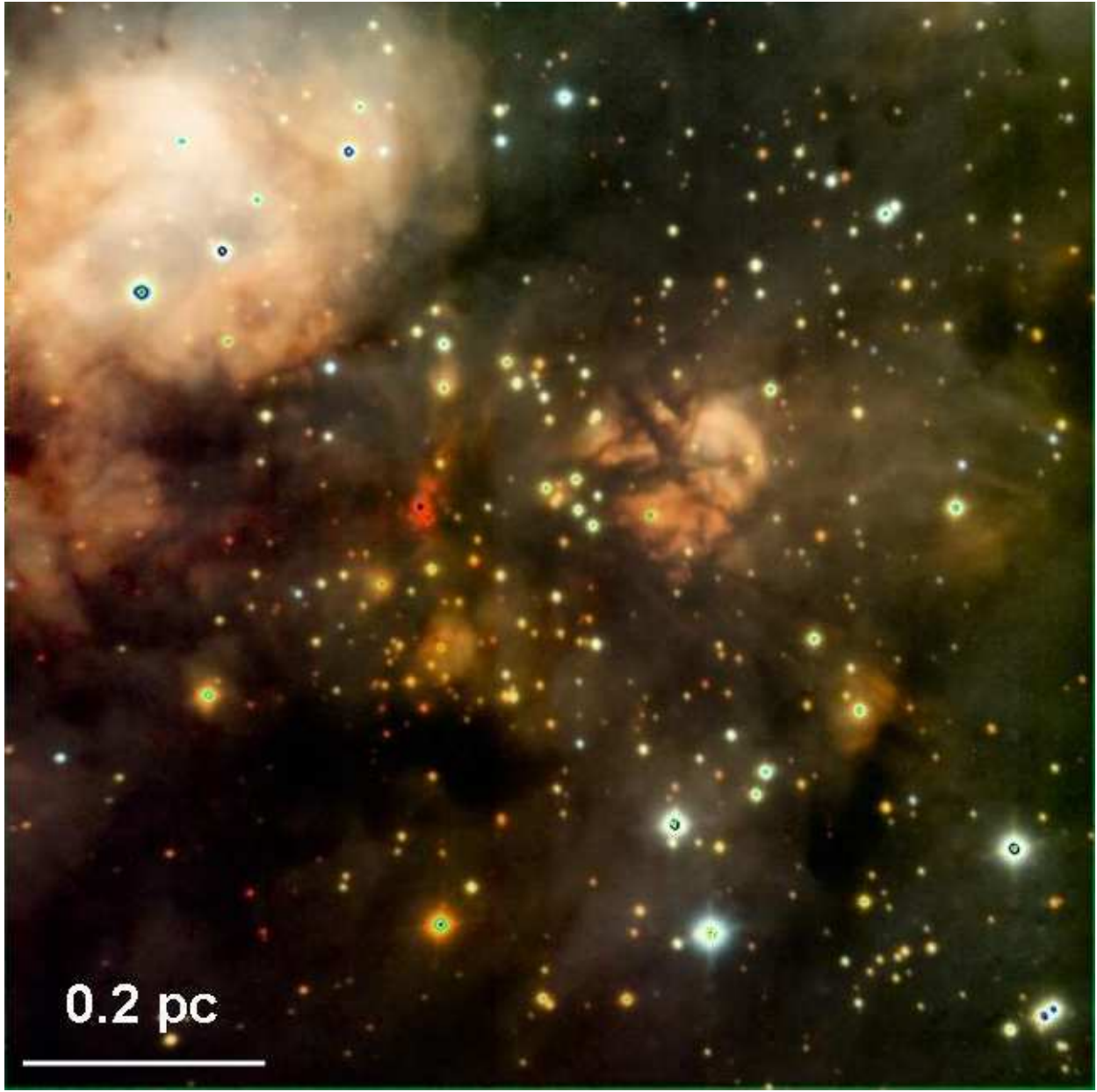}
\caption{$JHK$ composite image of the W3 Main star-forming region 
($J: blue; H: green; K: red$) obtained with the CISCO
mounted on the Subaru 8.2 m telescope. The FOV is
$\sim$ 1\arcmin.8 $\times$ 1\arcmin.8. North is up, and east is to the left.
\label{fig2}}
\end{figure}

\clearpage

\begin{figure}
\epsscale{1}
\plottwo{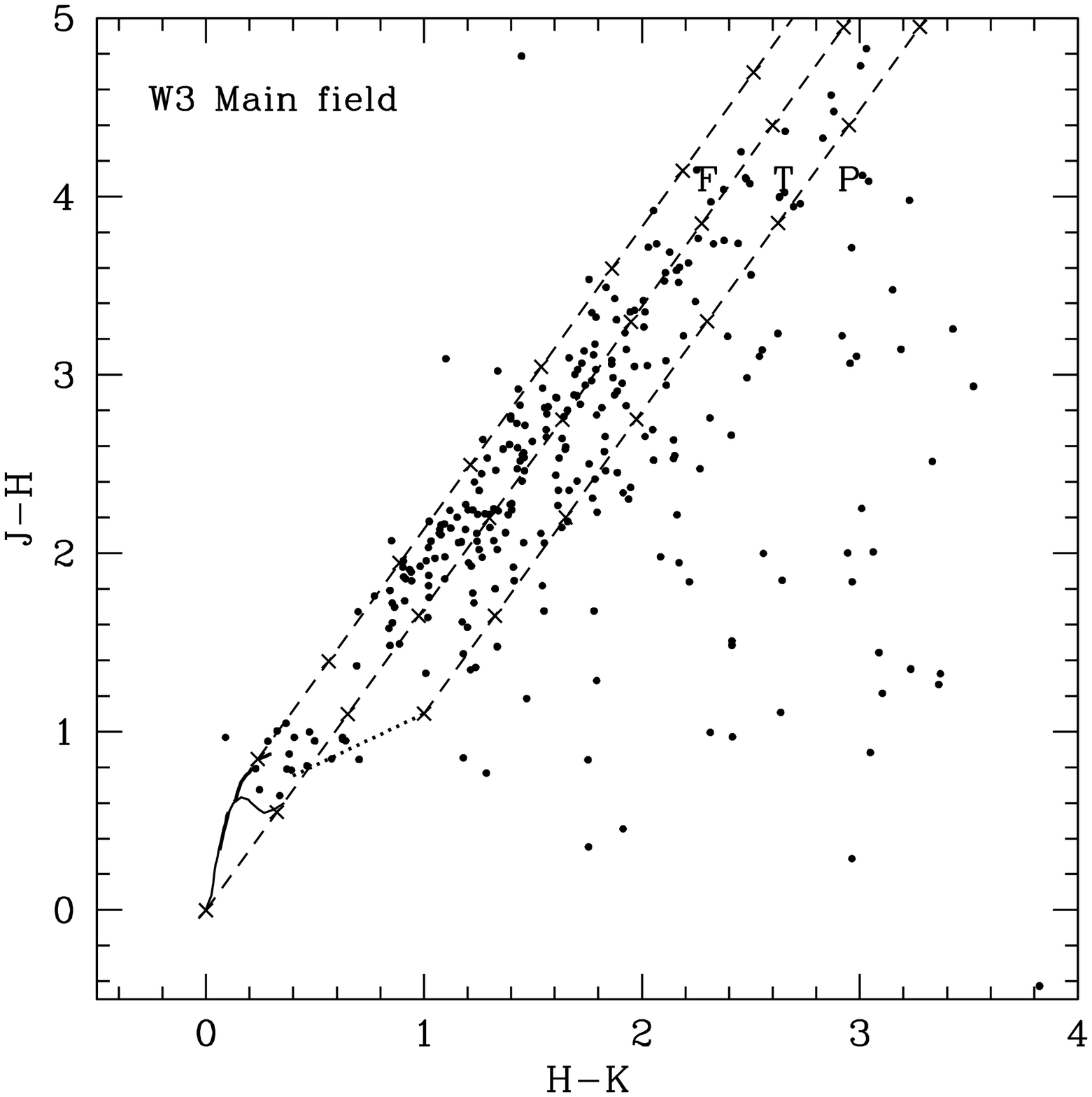}{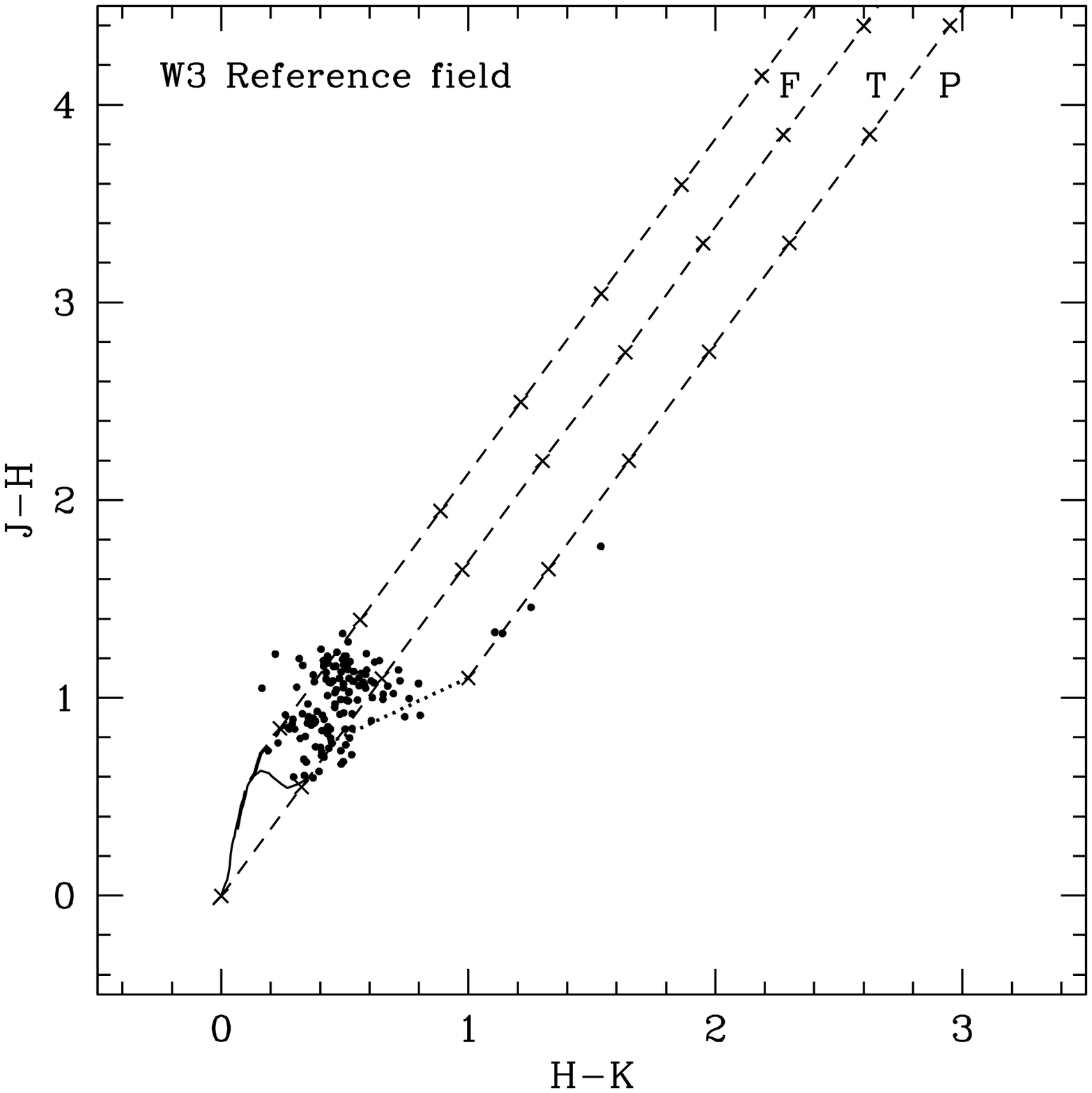}
\caption{CC diagram of the W3 Main star-forming region ({\it left}) and the 
reference field ({\it right}) for the sources
detected in $JHK$ bands with photometric errors less than 0.1 mag.
The sequences for field dwarfs ({\it solid curve}) and giants 
({\it thick dashed curve}) are from Bessell \& Brett (1988). The dotted 
line represents the locus of T Tauri stars (Meyer et al. 1997). Dashed 
straight lines represent the reddening vectors (Cohen et al. 1981). 
The crosses on the dashed lines are separated by $A_V$ = 5 mag. The plot
is classified into three regions, namely, F, T, and P for different classes 
of sources (see text for details). 
\label{fig3}}
\end{figure}

\clearpage

\begin{figure}
\plottwo{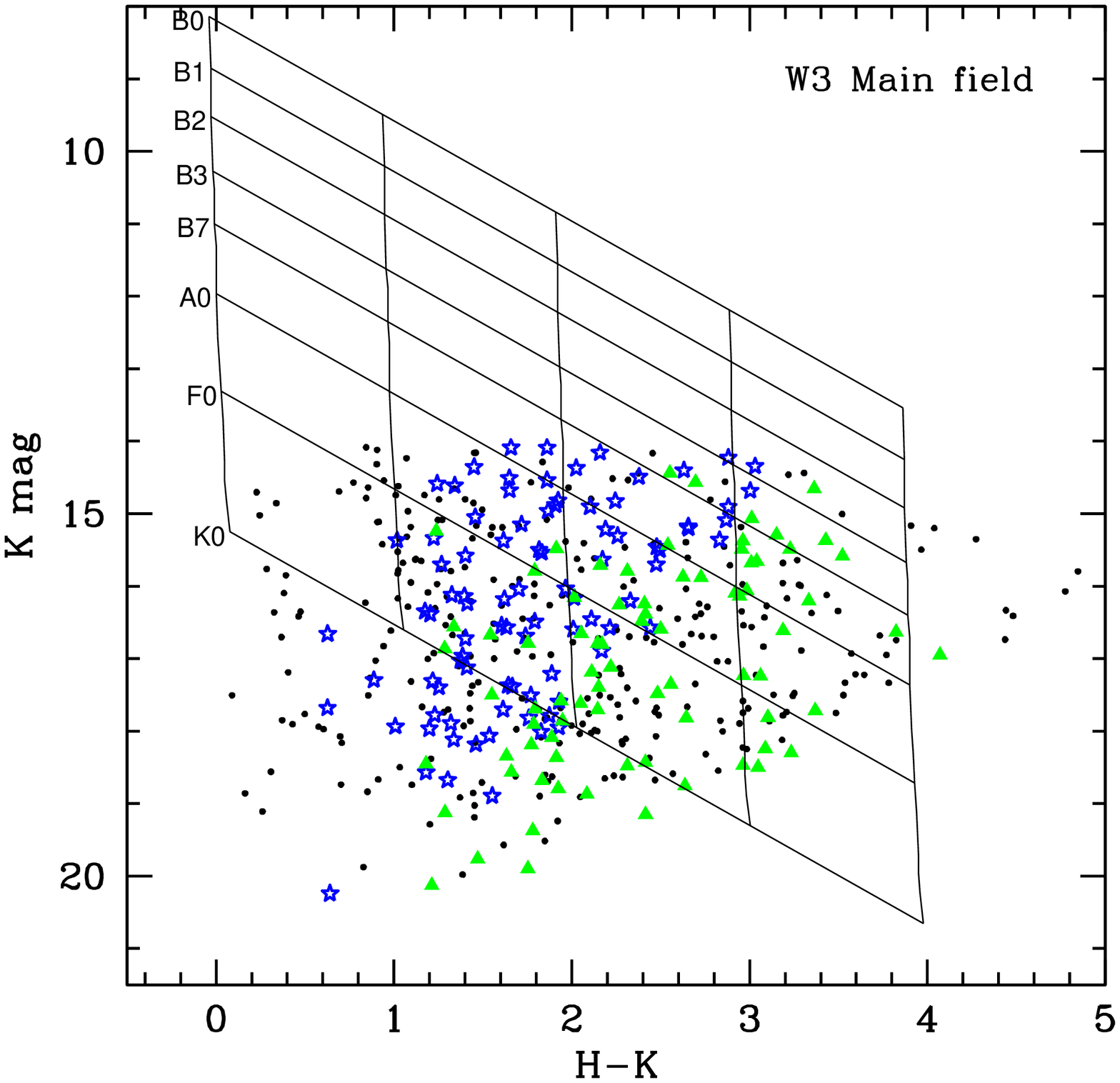}{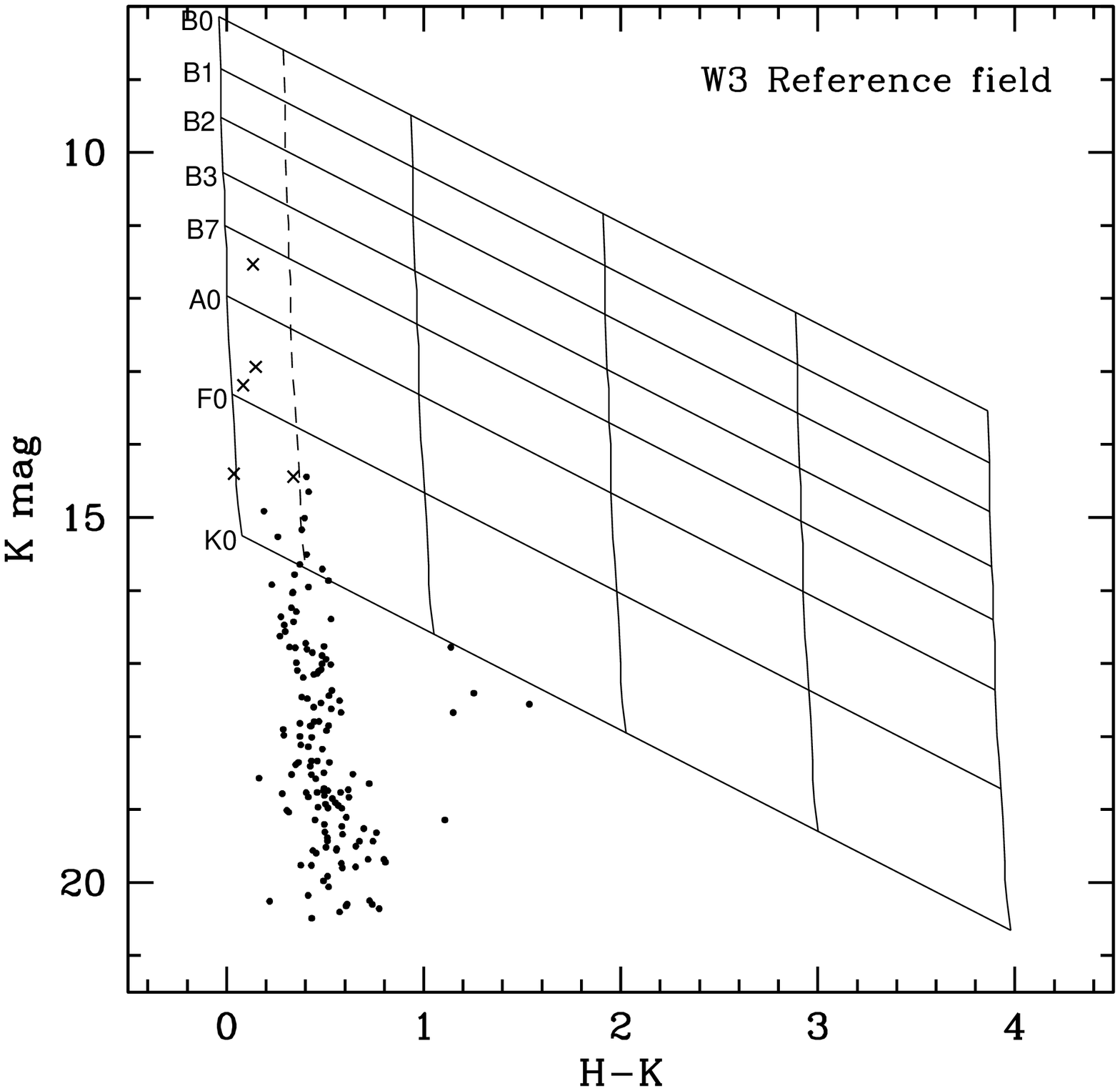}
\caption{$H-K/K$ CM diagram for the sources detected in
$H$ and $K$ bands with photometric errors less than 0.1 mag in the  W3 Main 
star-forming region ({\it left}) and the reference field ({\it right}).
Stars and filled triangles represent the YSOs identified from
the regions ``T'' and ``P'' in Fig. 3, respectively.
The vertical solid lines from left to right indicate the
main-sequence track at 1.83 kpc reddened by $A_V$ = 0, 15, 30, 45, and 60
mag, respectively. The intrinsic colors are taken from Koornneef (1983).
The dashed line in the reference field ({\it right}) shows the
main-sequence track reddened by $A_V$ = 5 mag, and the cross symbols denote
2MASS sources brigher than $K$= 14.5 mag, which are saturated in our
observations. Slanting
horizontal lines identify the reddening vectors (Cohen et al. 1981). 
\label{fig4}}
\end{figure}

\clearpage

\begin{figure}
\plotone{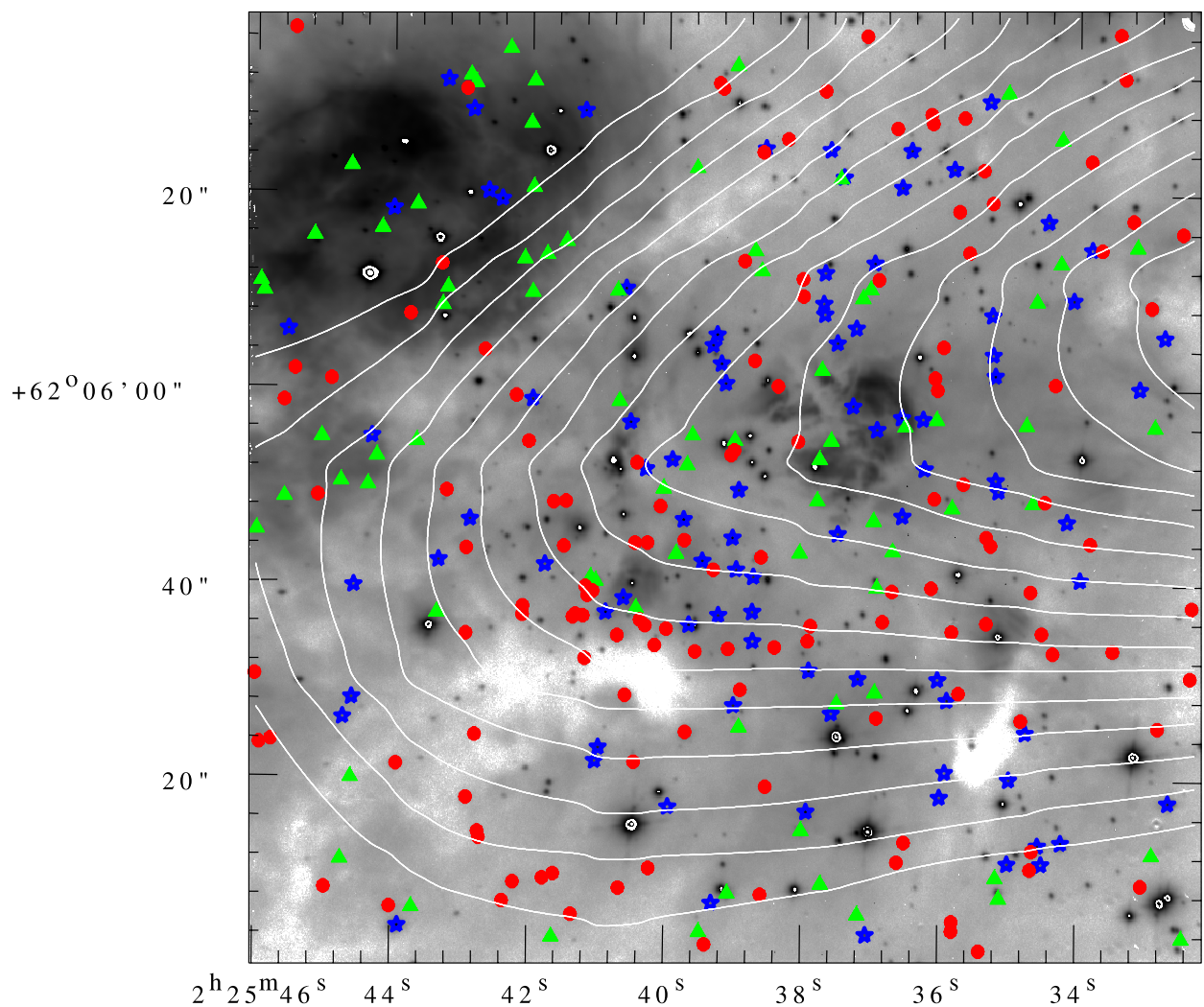}
\caption{Spatial distribution of the YSO color-excess candidates superposed on
the $K$-band image with a logarithmic intensity scale. The white contours
show the integrated emission of the $^{12}$CO line. The CO contours start
at 21 K km s$^{-1}$ and increase in steps of 2 K km s$^{-1}$. Blue stars
represent T Tauri and related sources (Class II), green triangles
indicate Class I sources, and red circles denote the
red sources ($H-K$ $>$ 2). The abscissa and the ordinate are in J2000.0 epoch.
\label{fig5}}
\end{figure}

\clearpage

\begin{figure}
\plotone{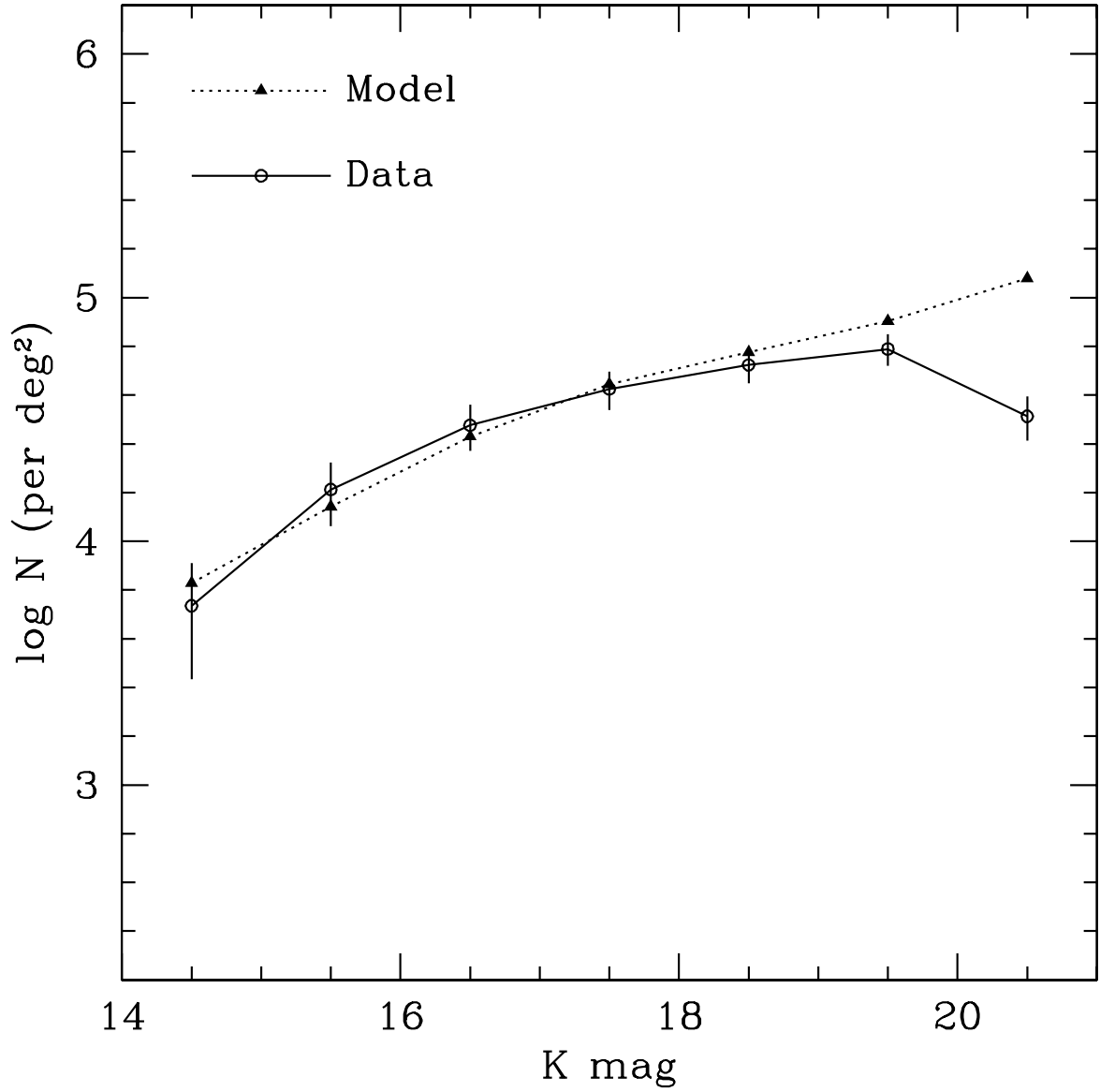}
\caption{Completeness-corrected KLF of the reference field (solid line) is 
compared with the model predicted KLF where an extinction of 
$A_V$ = 5 mag is applied to all the model stars (dotted line). The KLF
slope ($\alpha$; see \S 4.4) of the dotted line (model simulation) is
0.20$\pm$0.02.
\label{fig6}}
\end{figure}

\begin{figure}
\plotone{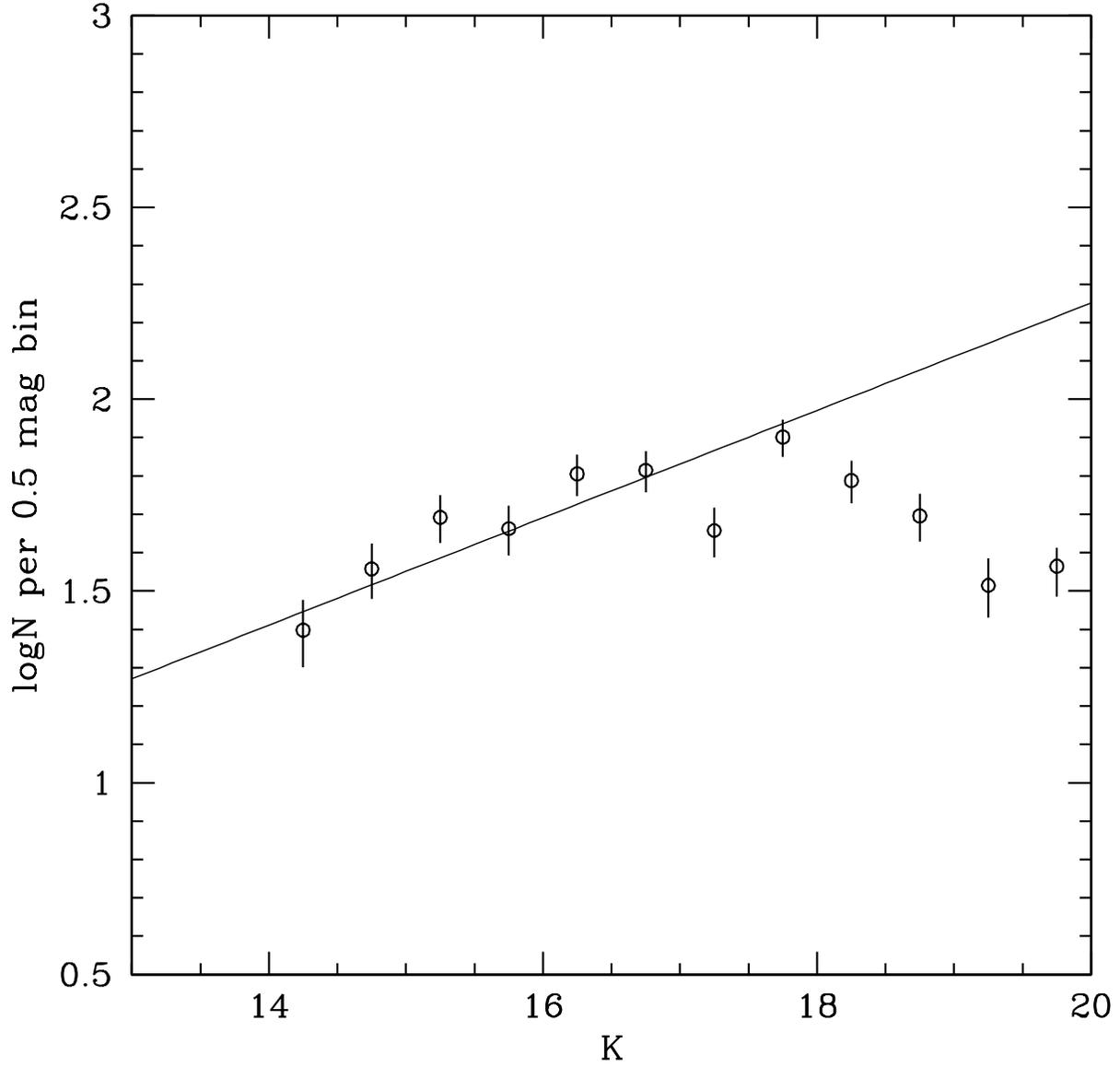}
\caption{Logarithm of the completeness-corrected and field star-subtracted 
KLF of the W3 Main region. The solid line is the least-square linear fit to 
the data points.
\label{fig7}}
\end{figure}

\clearpage

\begin{figure}
\plotone{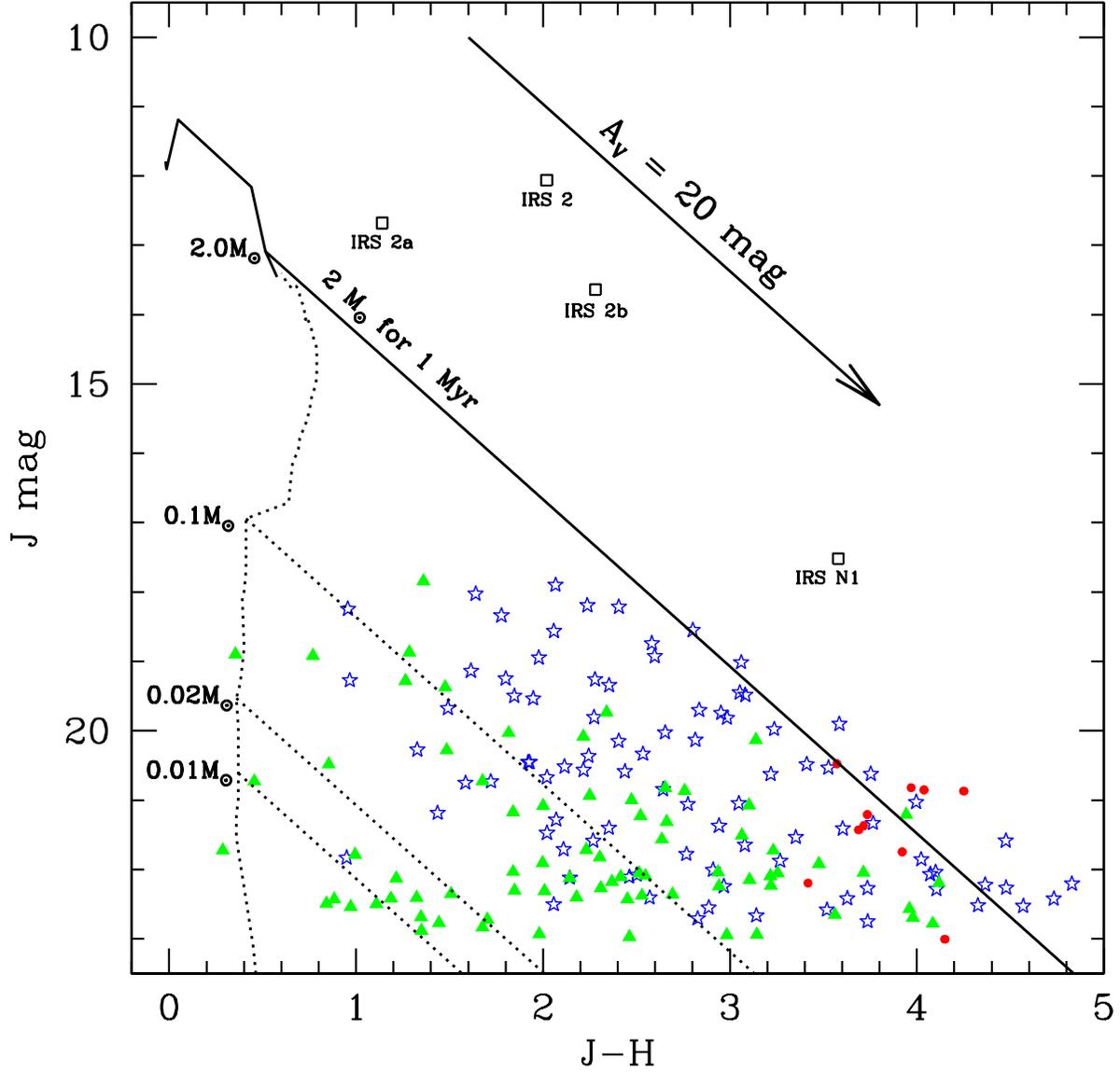}
\caption{CM diagram for the YSO candidates in W3 Main.
Class II candidates are indicated by stars, filled triangles represent Class I
candidates, and the filled circles show red sources with $H - K >$ 2 with
$J$-band counterparts. The solid and dotted curves denote the loci of 
1 Myr old PMS stars; derived from the models of Palla \& Stahler (1999), 
and Baraffe et al. (1998; 2003), respectively. Masses range from 
1.4 to 4.0 M$_{\odot}$ and 0.0005 to 1.4 M$_{\odot}$ from bottom to top, for 
the solid and dotted curves, respectively. Also shown are the 
positions of known IRS sources (see Table 1 in Ojha et al. 2004a).
\label{fig8}}
\end{figure}

\clearpage

\begin{figure}
\plottwo{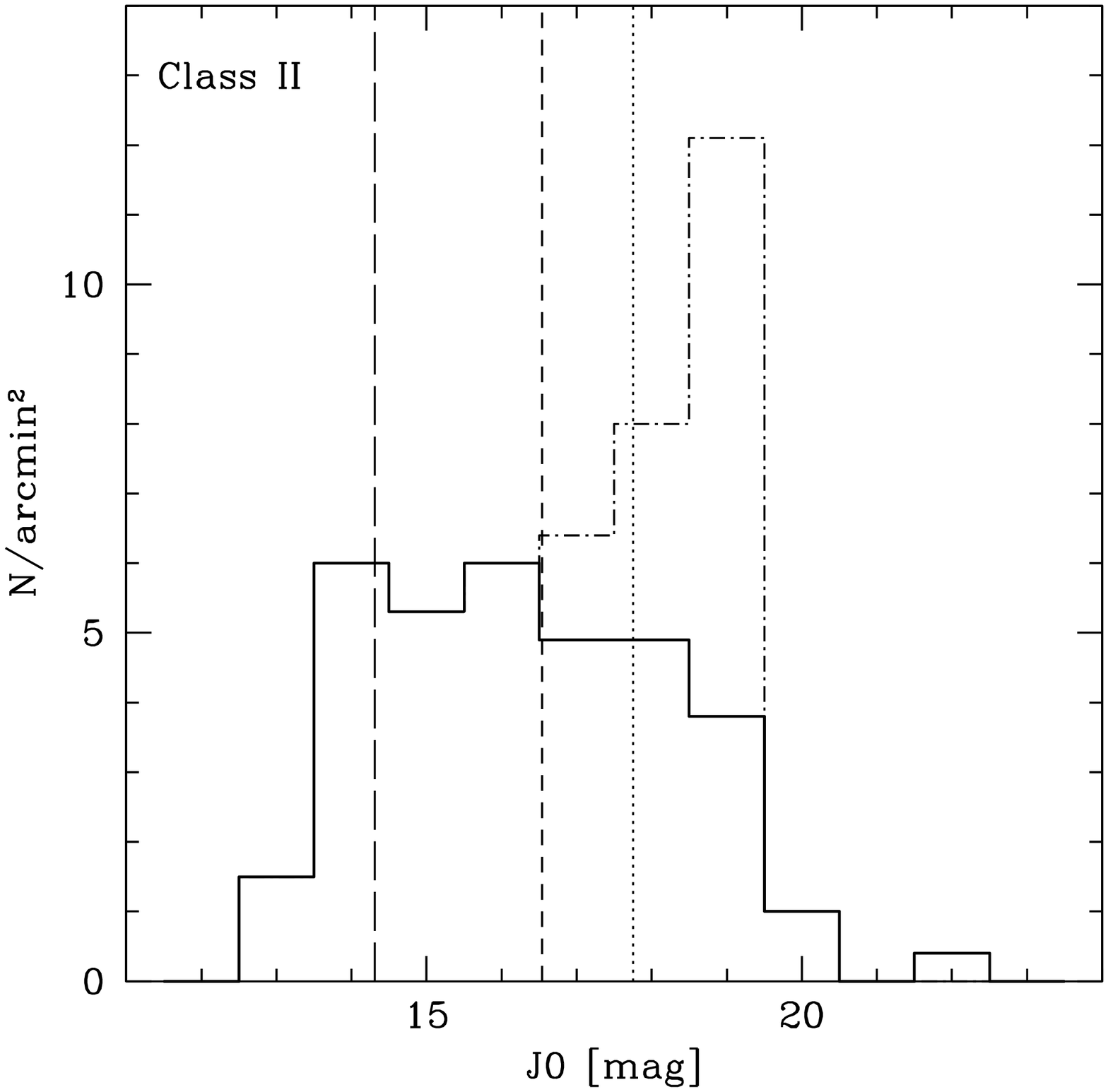}{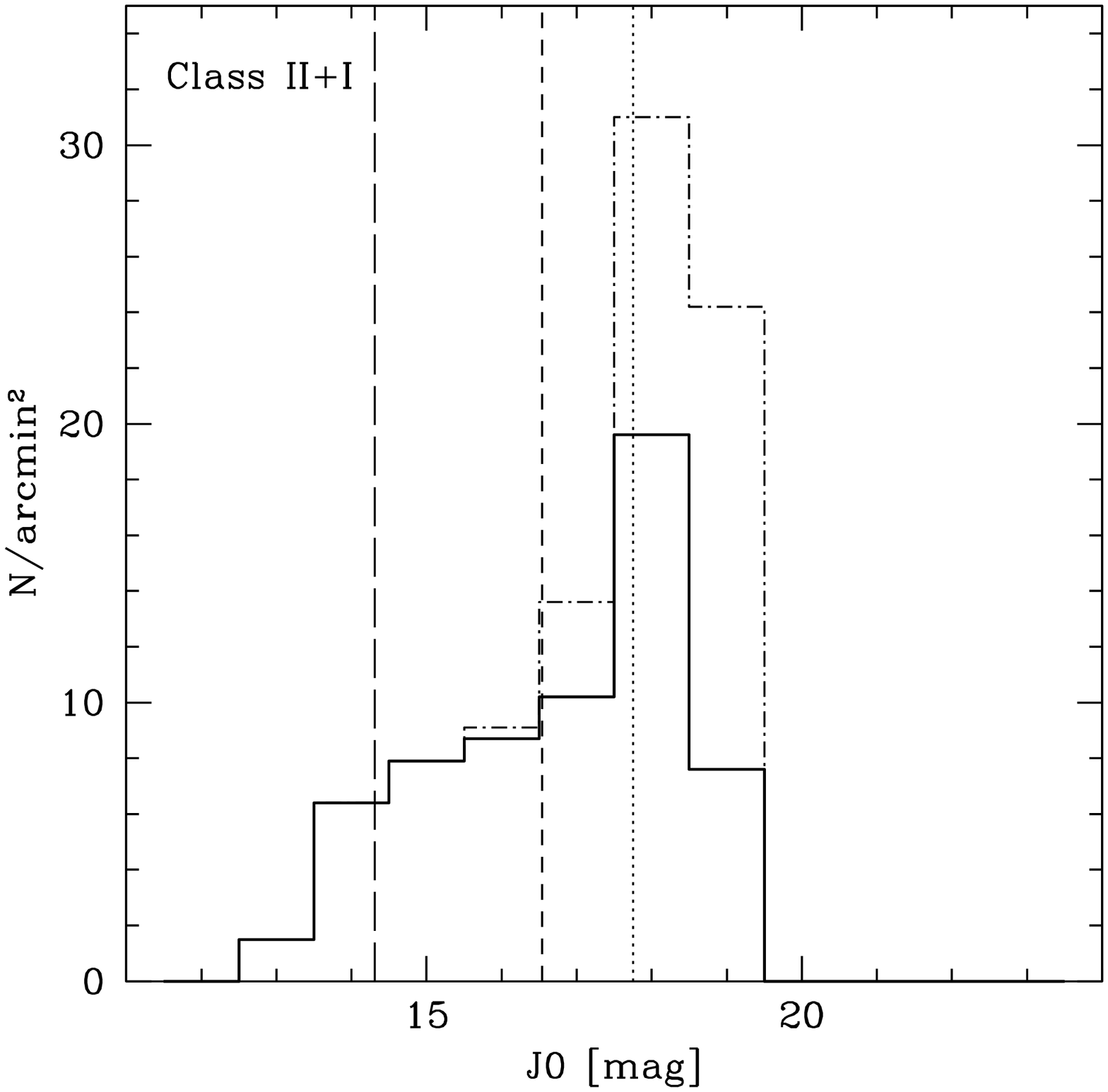}
\caption{Reddening-corrected $J$-band luminosity functions (solid line) of the 
identified Class II ({\it left}) and combined (Class II + I) objects 
({\it right}). The vertical dotted line represents the 
star-BD boundary at 1 Myr, 
while the short-dashed and long-dashed lines represent 90\% and 100\% 
completeness limits, respectively. The dashed-dotted line indicates the
JLF corrected for the completeness in the respective $J$-band magnitude bins.   
\label{fig9}}
\end{figure}

\clearpage

\begin{figure}
\plotone{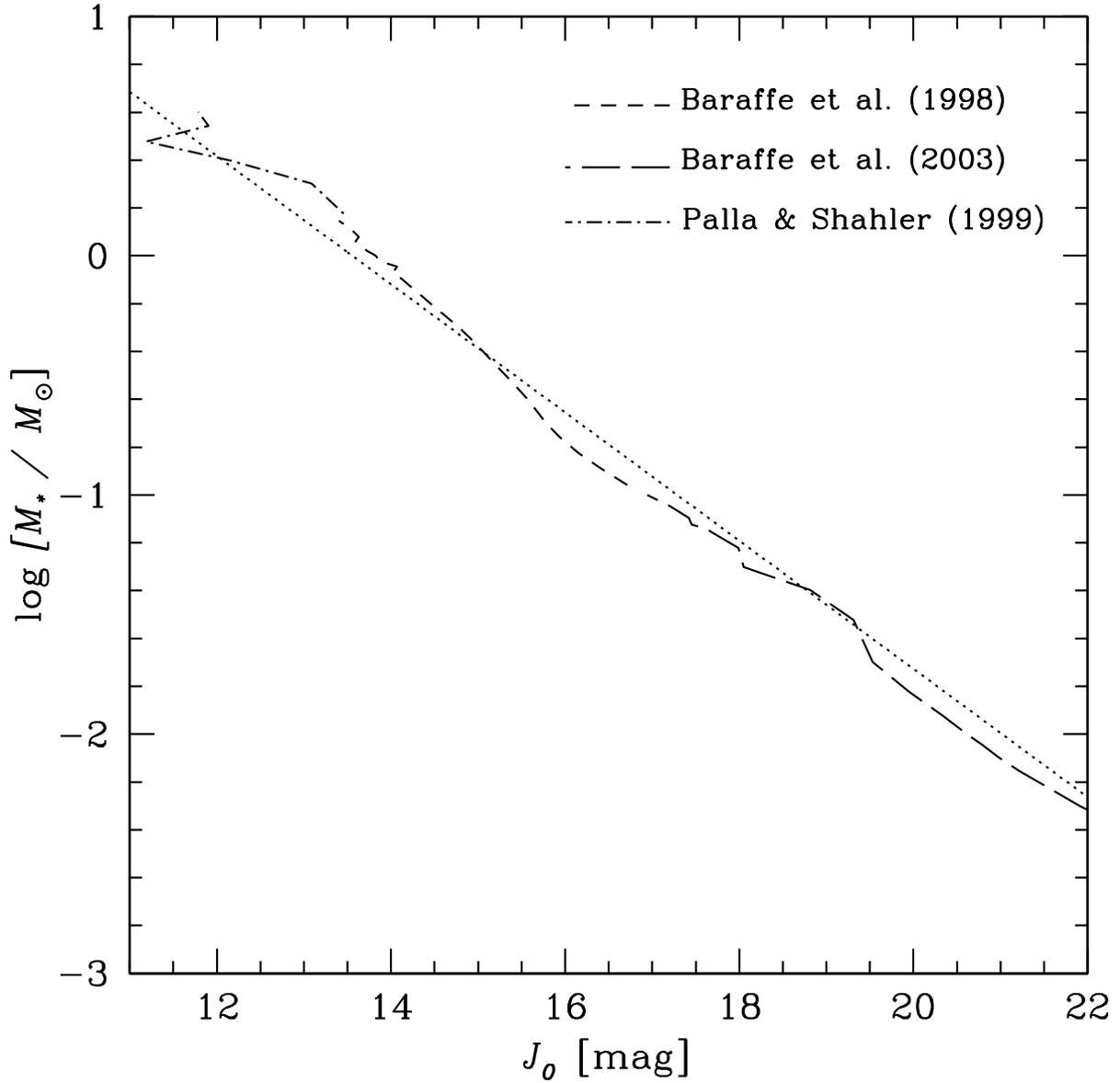}
\caption{$J$-band luminosity -- mass relationship for low-mass YSOs with an
age of 1 Myr based on the evolutionary models of Baraffe et al. (1998; 2003),
and Palla \& Stahler (1999). The dotted line shows a least-square linear
fit to the mass range 0.0005 $<$ $M/M_{\rm \odot}$ $<$ 4.0.
\label{fig10}}
\end{figure}

\clearpage

\begin{figure}
\plottwo{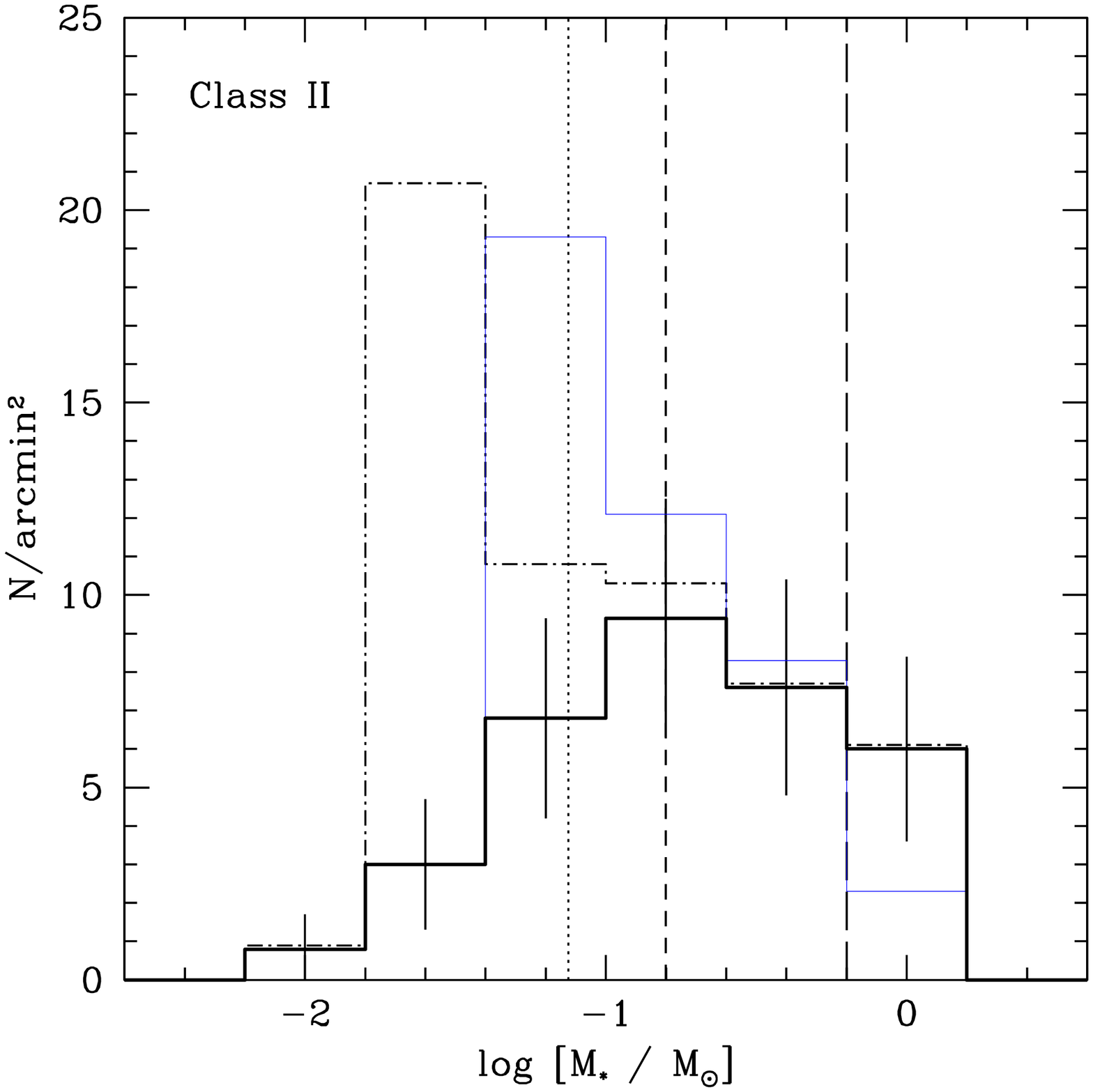}{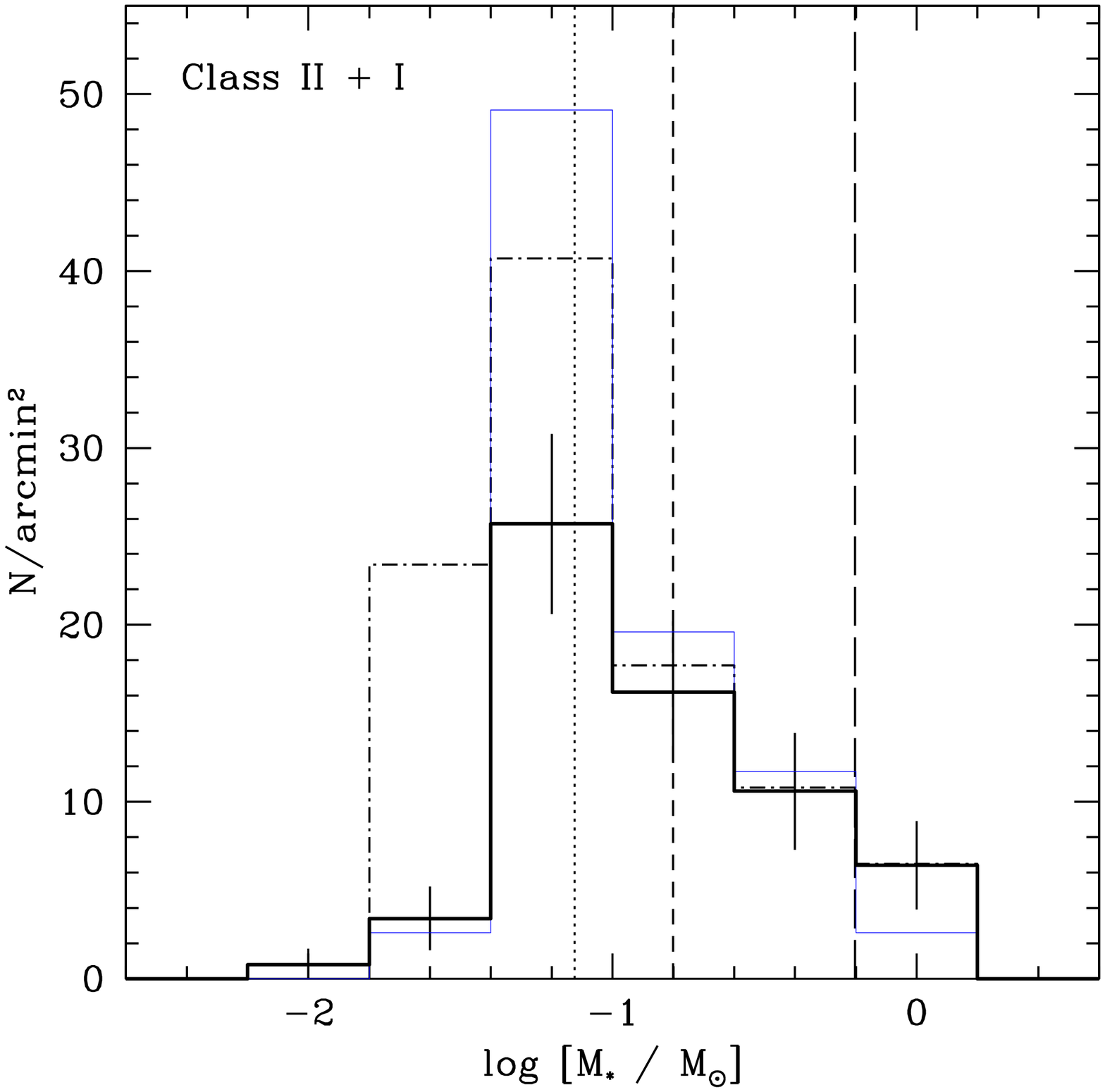}
\caption{Mass functions (with $\pm\sqrt{N}$ error bars) for the identified 
Class II ({\it left}) and 
combined (Class II + I) objects ({\it right}), based on the 
evolutionary models of Baraffe et al. (1998; 2003), and Palla \& Stahler 
(1999), for the age assumption of 1 Myr. The dashed-dotted line shows
the MF corrected for the completeness in the respective mass bins. 
The dotted, short-dashed, and long-dashed lines are the same as in Figure 9.
The thin continuous line indicates the completeness-corrected MF 
derived using reddening-corrected JLF with an average extinction
subtraction (see \S 4.6). 
\label{fig11}}
\end{figure}

\begin{figure}
\plottwo{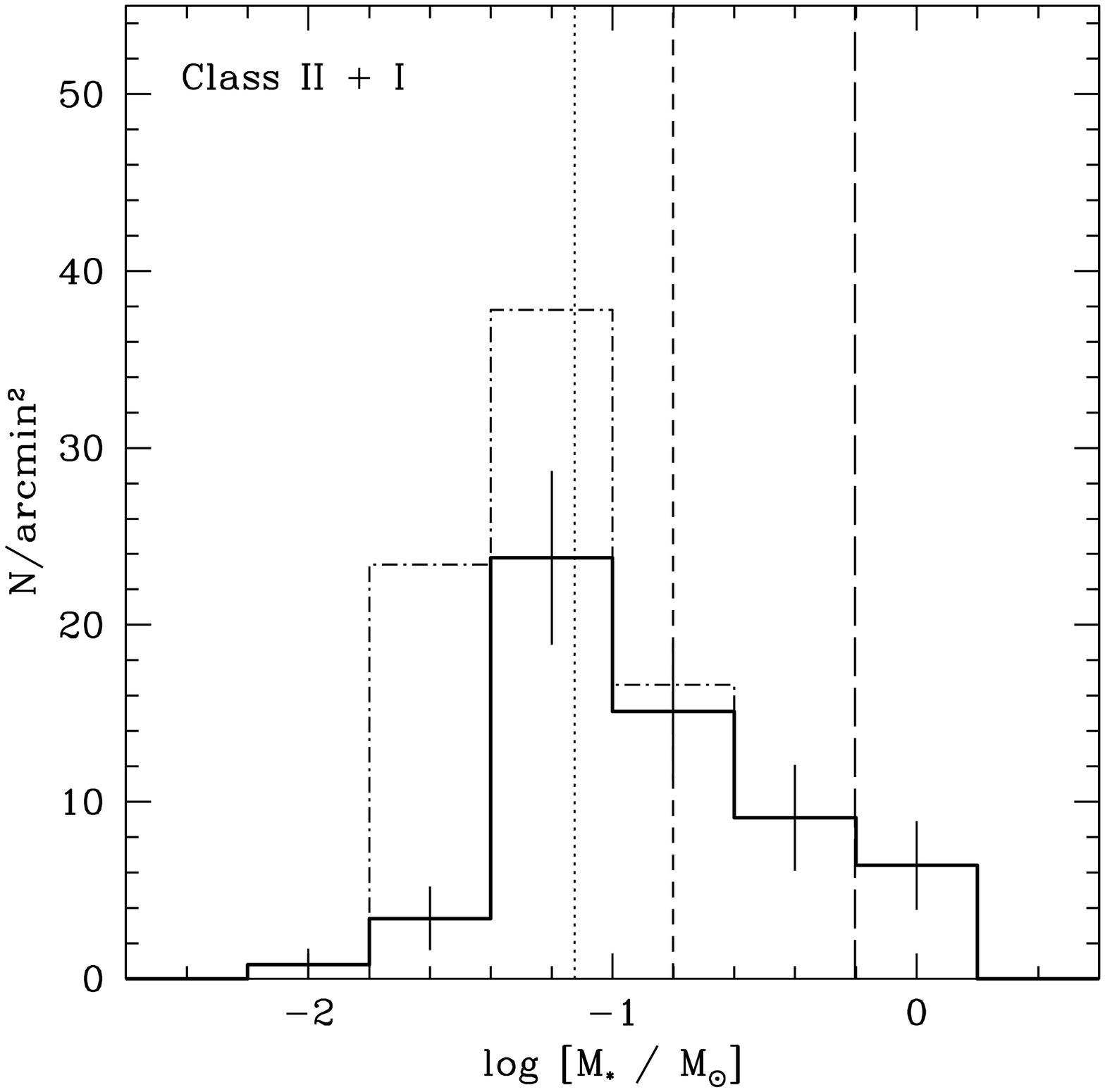}{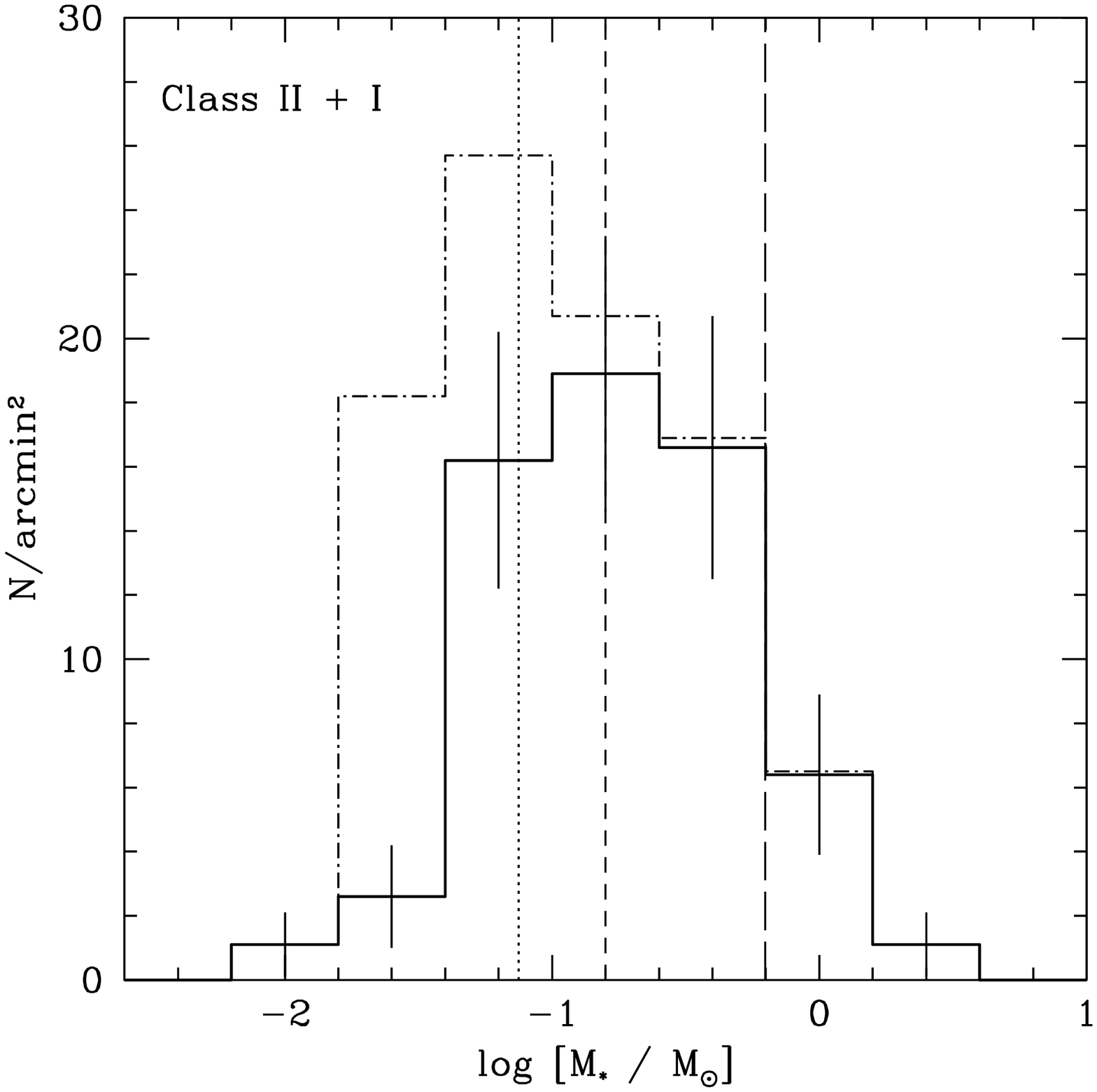}
\caption{({\it left}) The composite MF (with $\pm\sqrt{N}$ error bars) for 
Class II and Class I sources (solid line), where sources in the ``P'' region 
that fall below $J-H$ = 1.0 are not included. ({\it right}) The 
MF for all the sources in ``T'' (Class II) and ``P'' (Class I) 
regions (solid line). The sources are dereddened by using the extinction map 
(see text for details). The dashed-dotted line in both the diagrams 
indicates the MF corrected for the completeness in the respective mass bins. 
The dotted, short-dashed, and long-dashed lines are the same as in Figure 9.
\label{fig12}}
\end{figure}

\clearpage

\begin{figure}
\plotone{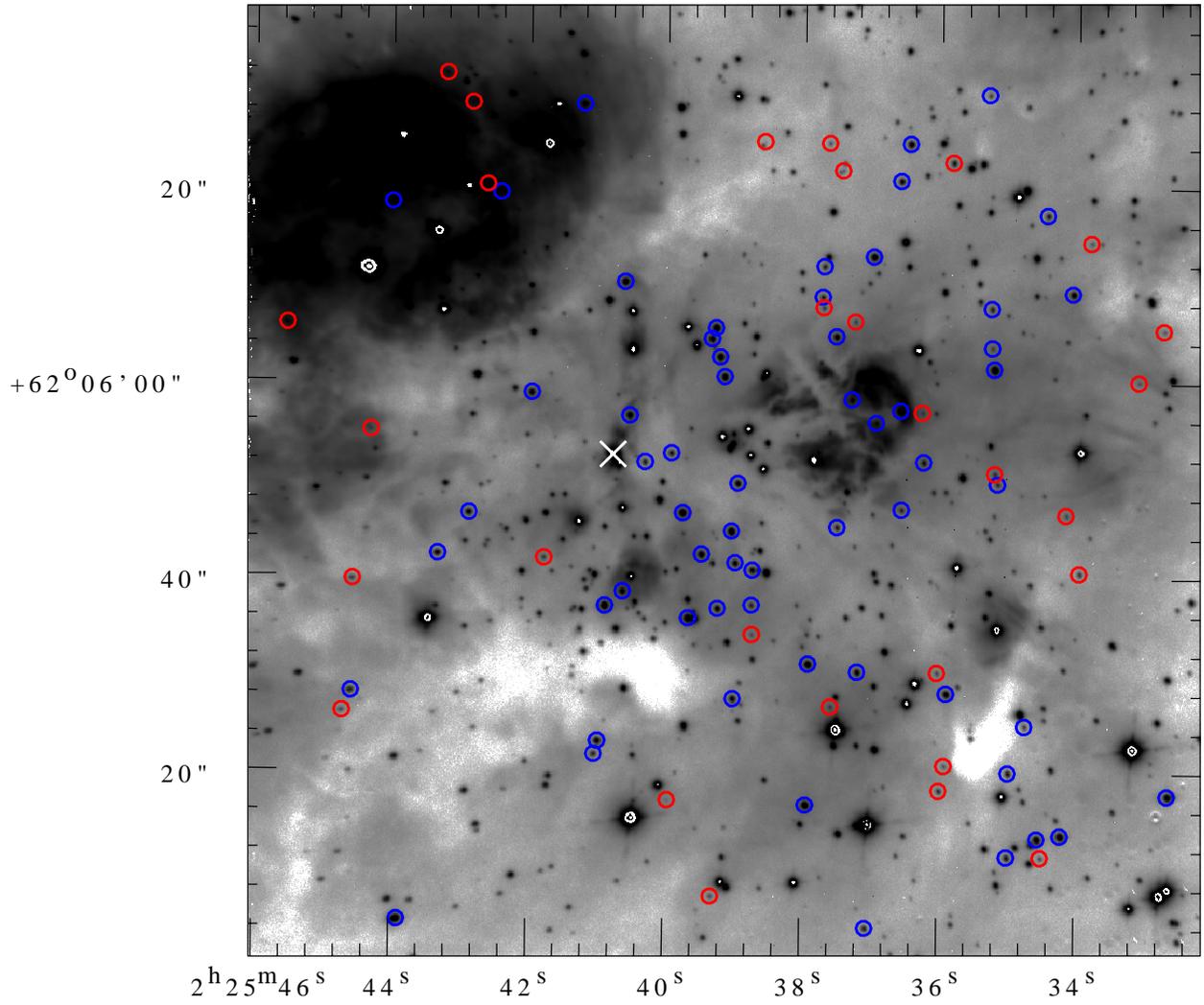}
\caption{Spatial distribution of the Class II sources in two magnitude
intervals (blue: $J_{\rm 0}$ $<$ 17.25 mag; red: $J_{\rm 0}$ $>$ 17.25 mag) 
superposed on the
$K$-band image. The center position of W3 IRS 5 is marked with a cross. 
\label{fig13}}
\end{figure}

\clearpage
\begin{figure}
\epsscale{0.8}
\plotone{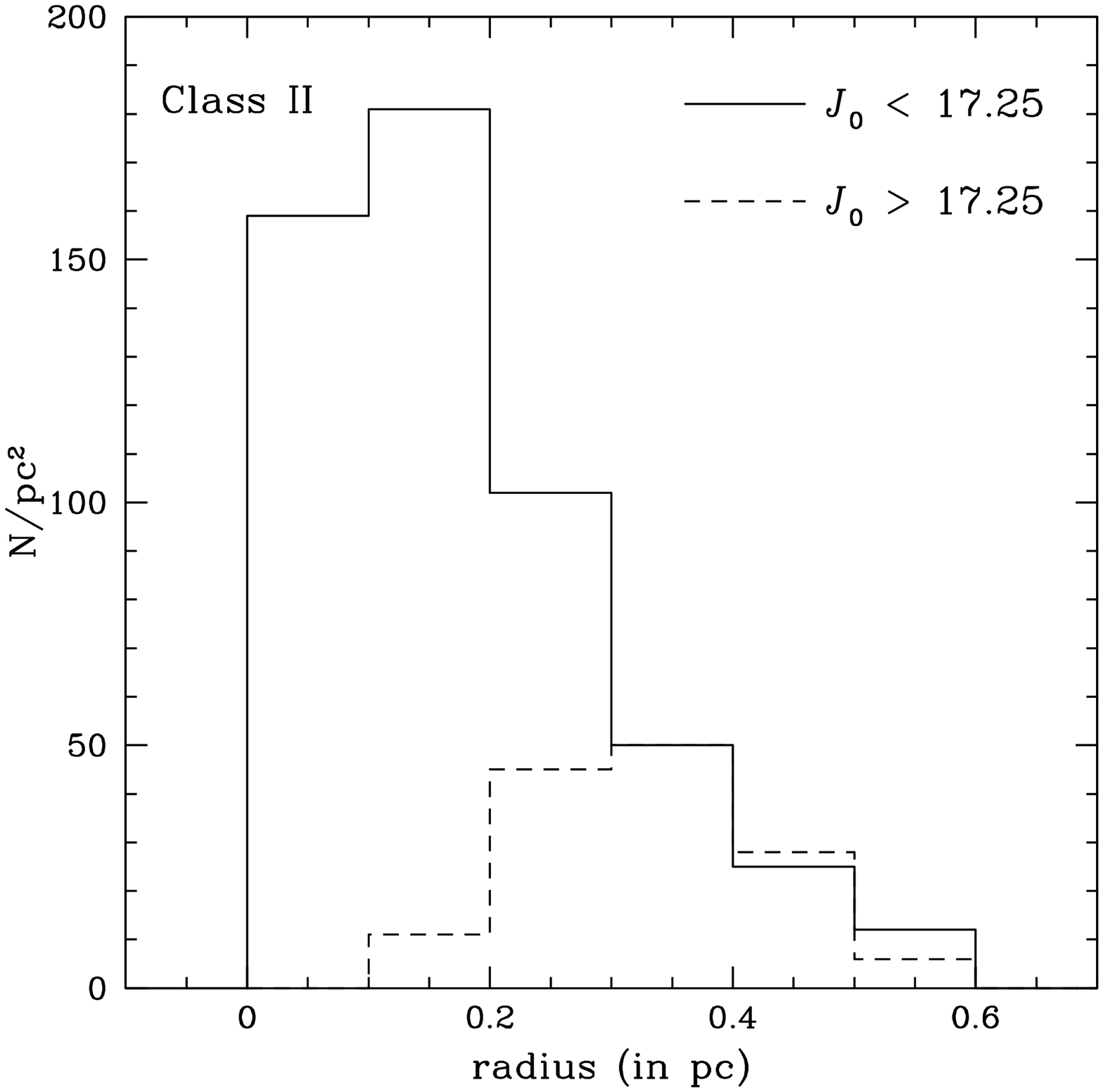}
\epsscale{1.0}
\plottwo{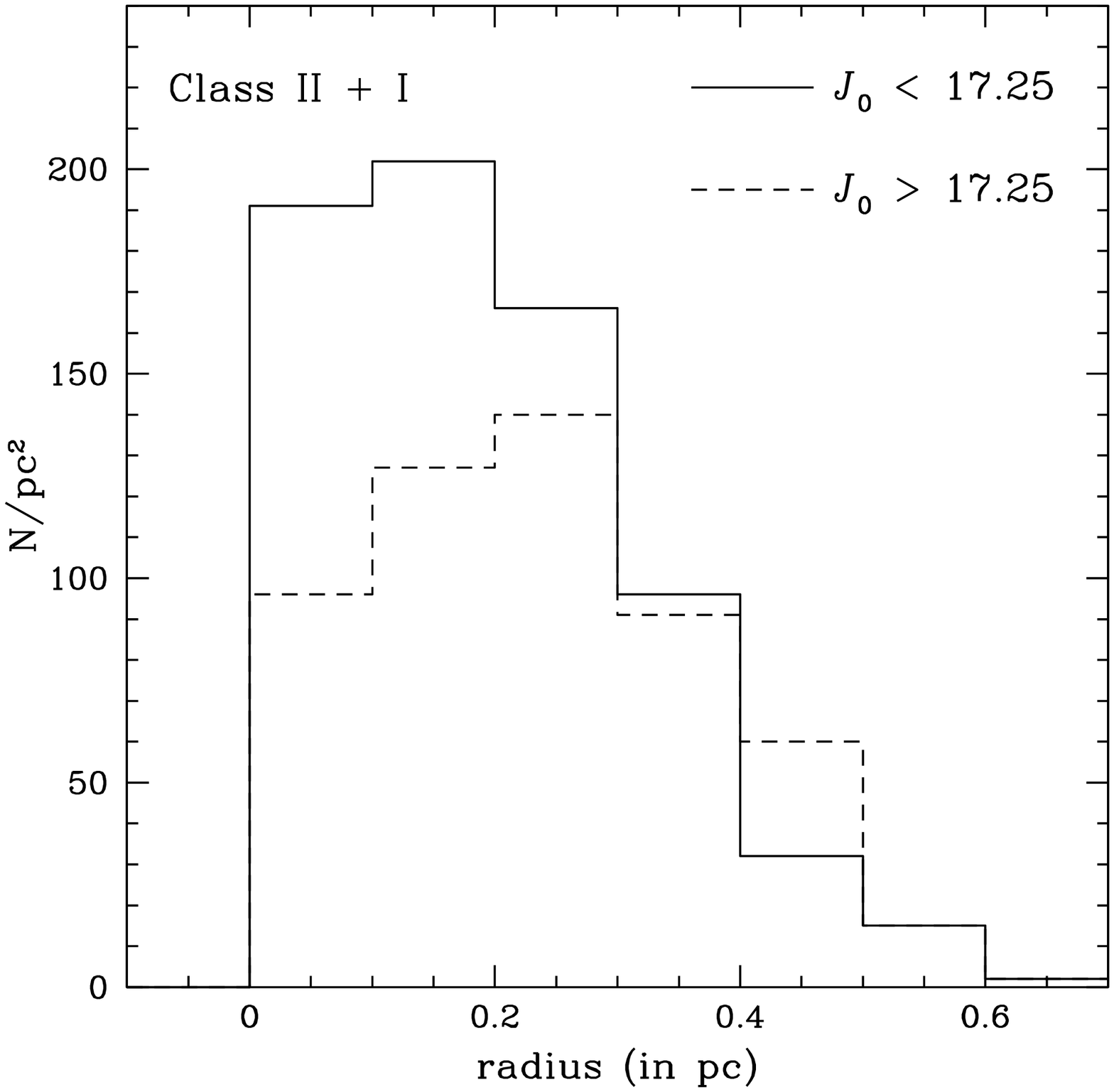}{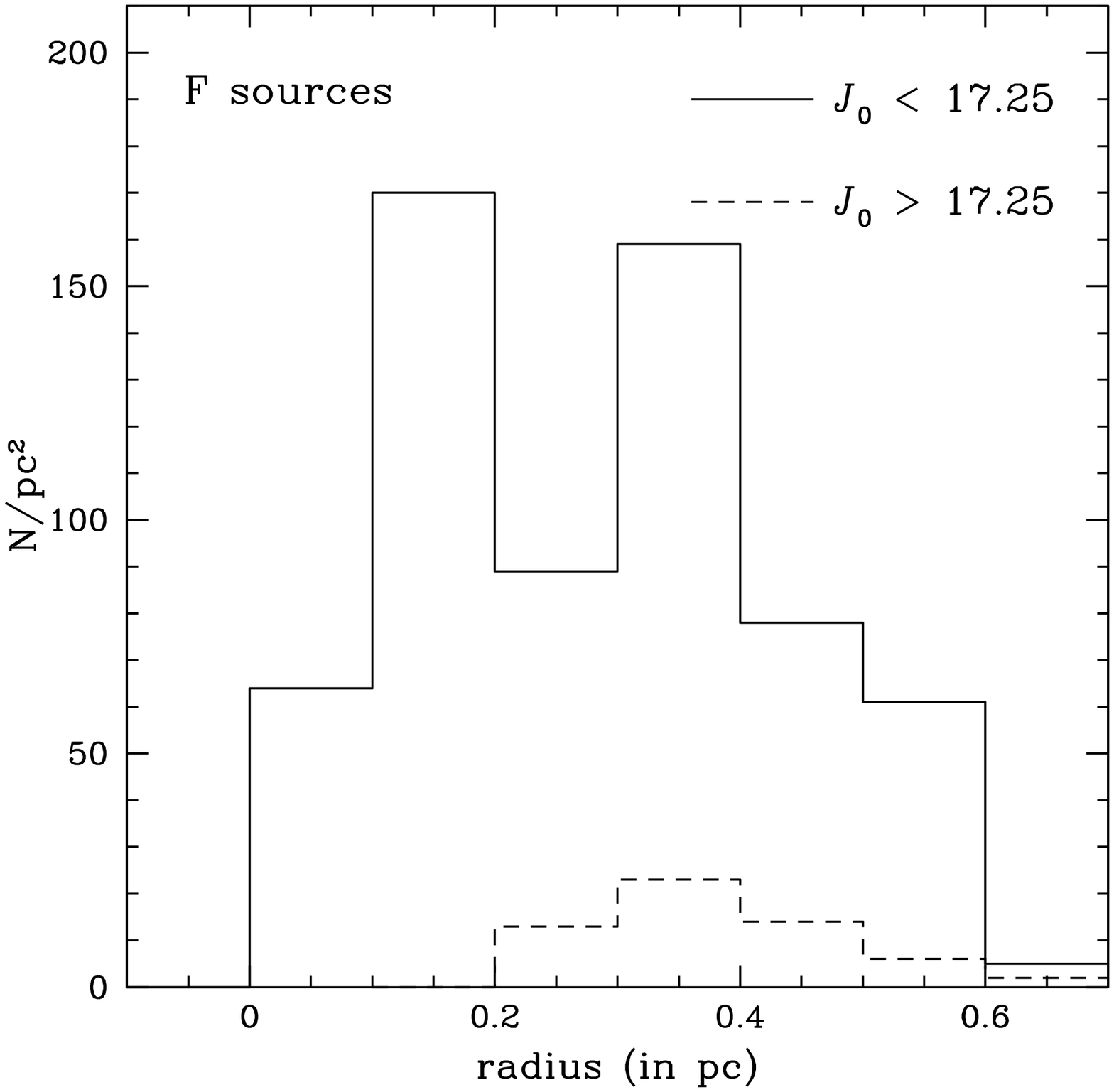}
\caption{The radial distribution of Class II ({\it top}), composite 
(Class II and I) ({\it bottom left}), 
and ``F'' sources ({\it bottom right}) in two dereddend $J$-band magnitude
intervals. The radial distance is calculated with respect to the 
positions of W3 IRS 5.  
\label{fig14}}
\end{figure}

\begin{figure}
\epsscale{1.0}
\plotone{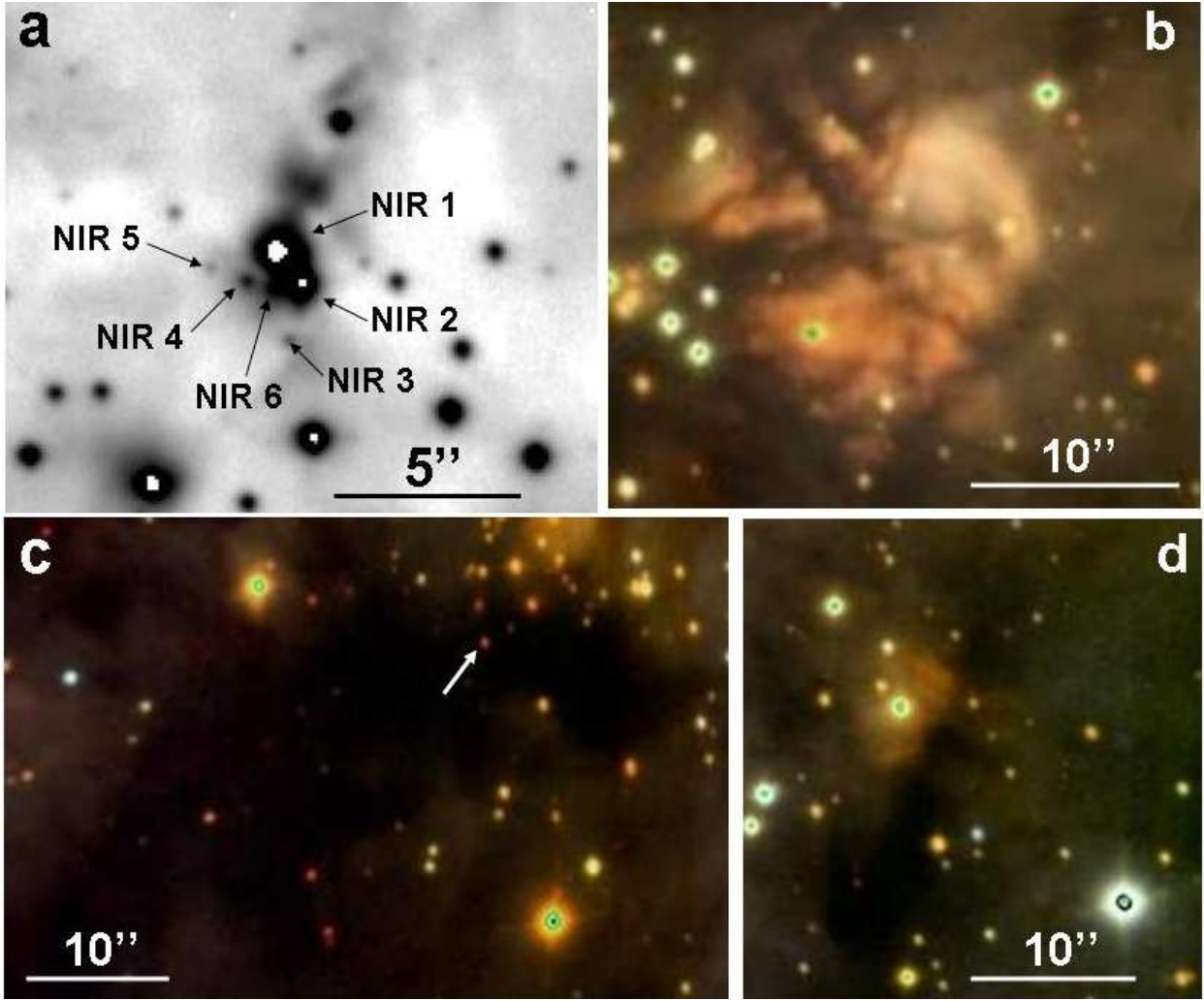}
\caption{Enlarged view of the $K$-band and color composite images of selected 
areas (see Figs. 1, 2, and the Appendix). (a) Section of the $K$-band 
image around W3 IRS 5 region and the neighboring red sources and 
nebulosities. In its immediate vicinity extremely red infrared sources,
are located (marked by NIR 3, NIR 4, NIR 5, and NIR 6; see \S 4.2).
(b) Fluffy diffuse nebula W3 B, probably illuminated by the bright
star(s) located southeast of this feature.  
(c) Dark filamentary lanes with irregular shapes seen throughout
the whole W3 Main star-forming region. An infrared 
source marked by an arrow, with large color excess, is located 
inside one of them. (d) A faint nebulosity detected around the
ultracompact H II region W3 E.
\label{fig15}}
\end{figure}

\end{document}